\documentclass[fleqn,usenatbib]{mnras}

\usepackage{newtxtext,newtxmath}
\usepackage[T1]{fontenc}
\usepackage{ae,aecompl}

\usepackage{graphicx}	
\usepackage{amsmath}	
\usepackage{amssymb}	
\usepackage[utf8]{inputenc}
\usepackage[T1]{fontenc}

\title[Standard Galactic Field RR Lyrae II]{Standard Galactic Field RR Lyrae II: A Gaia DR2 calibration of the period-Wesenheit-metallicity relation}

\author[J. R. Neeley et al.]{
Jillian R. Neeley,$^{1}$\thanks{E-mail: neeleyj@fau.edu}
Massimo Marengo,$^{2}$
Wendy L. Freedman,$^{3}$
Barry F. Madore,$^{4}$ \newauthor
Rachael L. Beaton,$^{4,5}$\thanks{Hubble Fellow}
Dylan Hatt,$^{3}$
Taylor Hoyt,$^{3}$
Andrew J. Monson,$^{6}$
Jeffrey A. Rich,$^{4}$ \newauthor
Ata Sarajedini,$^{1}$
Mark Seibert,$^{4}$
and Victoria Scowcroft$^{7}$
\\
$^{1}$Department of Physics, Florida Atlantic University, 777 Glades Rd, Boca Raton, FL 33431\\
$^{2}$Department of Physics \& Astronomy, Iowa State University, Ames IA 50011\\
$^{3}$Department of Astronomy \& Astrophysics, University of Chicago, 5640 South Ellis Avenue, Chicago IL 60637\\
$^{4}$The Observatories of the Carnegie Institution for Science, 813 Santa Barbara Street, Pasadena CA 91101\\
$^{5}$Department of Astrophysical Sciences, Princeton University, 4 Ivy Lane, Princeton, NJ~08544\\
$^{6}$Department of Astronomy and Astrophysics, The Pennsylvania State University, 525 Davey Lab, University Park, PA 16802, USA\\
$^{7}$Department of Physics, University of Bath, Claverton Down, Bath, BA2 7AY, UK
}

\date{Accepted 2019 October 2. Received 2019 October 1; in original form 2019 May 6}

\pubyear{2019}

\begin{document}
\label{firstpage}
\pagerange{\pageref{firstpage}--\pageref{lastpage}}
\maketitle

\begin{abstract}
RR Lyrae stars have long been popular standard candles, but significant advances in methodology and technology have been made in recent years to increase their precision as distance indicators.  We present multi-wavelength (optical $UBVR_cI_c$ and \emph{Gaia} $G, BP, RP$; near-infrared $JHK_s$; mid-infrared $[3.6], [4.5]$) period-luminosity-metallicity (PLZ), period-Wesenheit-metallicity (PWZ) relations, calibrated using photometry obtained from The Carnegie RR Lyrae Program and parallaxes from the \emph{Gaia} second data release for 55 Galactic field RR Lyrae stars. The metallicity slope, which has long been predicted by theoretical relations, can now be measured in all passbands. The scatter in the PLZ relations is on the order of 0.2 mag, and is still dominated by uncertainties in the parallaxes. As a consistency check of our PLZ relations, we also measure the distance modulus to the globular cluster M4, the Large Magellanic Cloud (LMC) and the Small Magellanic Cloud (SMC), and our results are in excellent agreement with estimates from previous studies. 
\end{abstract}




\section{Introduction} \label{sec:intro}

RR Lyrae variabels (RRL) have a long history as standard candles, beginning in the 1950s \protect\citep[see][for a review]{smith_rr_1995}.  RRL are evolved, low-mass stars, and therefore have lower luminosities than Cepheid variables, limiting their reach as distance indicators. However, given their lower masses, and therefore larger relative numbers (through the initial mass function), the number of available RRL is at least an order of magnitude greater than Cepheids, offering a more detailed view and the possibility to study the structure of galaxies \protect\citep{pietrukowicz_deciphering_2015, sesar_machine-learned_2017, kunder_impact_2018}. Additionally, RRL allow us the potential to probe regions without active star formation, such as elliptical galaxies, which lack Classical Cepheids, and whose distances cannot be measured with the traditional distance scale. 

Beyond geometric techniques, RRL distances are primarily estimated through two methods. In the optical, a visual band luminosity-metallicity (LZ) relation is used, first calibrated by \protect\citep{sandage_evidence_1981, sandage_oosterhoff_1981}. Over the years this relation has been refined, and is still actively used today \protect\citep[][and references therein]{chaboyer_globular_1996, cacciari_globular_2003}. In the infrared, similar to the Leavitt Law for classical Cepheid variables, RRL obey a period-luminosity (PL) relation, first discovered in the $K$ band \protect\citep{longmore_rr_1986}. Infrared PL relations offer greater precision given their smaller scatter, lower dependence on extinction uncertainties, and reduced evolutionary effects \protect\citep{bono_theoretical_2001, dallora_rr_2006}. They have been consistently employed in the field \protect\citep[e.g.][]{dambis_rr_2013, muraveva_rr_2018}, in globular clusters \protect\citep[e.g.][]{coppola_distance_2011, sollima_rr_2006, braga_distance_2015}, and in nearby galaxies \protect\citep[e.g.][]{moretti_vmc_2014, karczmarek_araucaria_2015, muraveva_vmc_2018}. With the rise of space telescopes operating at infrared wavelengths in recent years, mid-infrared PL relations have also become popular distance tools for RRL \protect\citep[e.g.][]{dambis_mid-infrared_2014, klein_mid-infrared_2014, neeley_distance_2015, muraveva_rr_2018}. The empirical studies are complemented by multi-band theoretical studies \protect\citep{catelan_rr_2004, marconi_new_2015, neeley_new_2017}. Additionally, RRL distance studies have continued to mirror procedures for Cepheids, employing (theoretically calibrated) period-Wesenheit (PW) relations in optical bands \protect\citep{monelli_variable_2018}. For a more detailed history of RRL as standard candles, we refer the reader to the recent review by \protect\citet{beaton_old-aged_2018}. 

The use of RRL as accurate distance indicators today is experiencing another revolutionary change because of the \emph{Gaia} mission \protect\citep{gaia_collaboration_gaia_2016, gaia_collaboration_gaia_2016-1, gaia_collaboration_gaia_2018-1}. Previously, very few RRL had measured parallaxes, and the use of RRL as standard candles relied on calibration from less certain methods. In the \emph{Hipparcos} catalog, only RR Lyrae itself had an uncertainty below 20\% \protect\citep{perryman_hipparcos_1997, van_leeuwen_validation_2007}. With the innovative use of the \emph{Hubble Space Telescope's} Fine Guidance sensor, \protect\citet{benedict_distance_2011} increased the number of galactic calibrators to five RRL. Now with the \emph{Gaia} mission, the number of RRL with an accurate distance has increased tremendously, and the potential of RRL is becoming a reality. As of the second data release (DR2), 140,784 RRL have been identified in the \emph{Gaia} catalog \protect\citep{clementini_gaia_2019}, 1840 of which have a parallax uncertainty below 10\%. This is a drastic improvement from the first data release, which included only 364 RRL with \emph{Tycho-Gaia Astrometric Solution} (TGAS) parallaxes \protect\citep{gaia_collaboration_gaia_2017}. There are, however, significant systematics (which are likely a function of magnitude, color, and sky position) that affect the data quality in DR2, and the uncertainties quoted in the catalog are likely underestimated \protect\citep{arenou_gaia_2018, lindegren_gaia_2018}. By the end of the \emph{Gaia} mission, we expect to have even more RRL in the database, a significant fraction of them with micro-arcsecond precision parallaxes \protect\citep{de_bruijne_gaia_2014}. 

The Carnegie RR Lyrae Program (CRRP) is poised to establish RRL as the foundation of a Population II distance scale, in addition to the tip of the red giant branch. This independent distance scale is an important check on the traditional Population I ladder, and will help to investigate the current tension in the measurement of $H_0$ \protect\citep{freedman_carnegie-chicago_2019, beaton_carnegie-chicago_2016, riess_milky_2018, planck_collaboration_planck_2018}. Using a sample of 55 nearby Galactic field RRL observed as part of CRRP, and presented in \protect\citet{monson_standard_2017}, hereafter Paper I, we derive new period-Wesenheit (PW) and period-Wesenheit-metallicity (PWZ) relations using data in 13 photometric bands. For the six longest bands ($I$ through $[4.5]$) we also provide PL and PLZ relations. The absolute magnitudes of these stars are determined using trigonometric parallaxes from \emph{Gaia's} DR2 \protect\citep{gaia_collaboration_gaia_2018, lindegren_gaia_2018}. Several recent studies have provided either PL or LZ relations for RRL using \emph{Gaia} parallaxes \protect\citep{sesar_probabilistic_2017, gaia_collaboration_gaia_2017, muraveva_rr_2018}.

Although our sample is relatively small, it offers a number of key advantages. First, we are using the same sample of stars to derive relations in a total of 13 photometric bands, from the optical to mid-infrared, while these previous works were limited to $V$, $K_S$, and WISE $W1$ alone. Furthermore, we have compiled well-sampled light curves in as many bands as possible, rather than adopting single-epoch magnitudes or arithmetic means of randomly sampled data. For the bands with few epochs available, we fit template light curves that are fine-tuned according to the data in other bands. In the mid-infrared bands, using \emph{Spitzer} data instead of WISE allows for greater photometric precision, and these bands are critical for future applications with the James Webb Space Telescope.  We also discuss the effects of a global zero-point offset on the parallaxes \protect\citep{arenou_gaia_2018} which, combined with the parallax uncertainties, limits the accuracy of our final relations. To test our PLZ relations for accuracy, we also derive distances to the globular cluster M4 and the Large and Small Magellanic Clouds, which are frequent targets for the calibration of standard candles because of their proximity and high accuracy distance determinations in the literature.

\begin{table*}
    \scriptsize
    \centering
    \caption{Gaia parameters}
    \label{tab:stars}
    \begin{tabular}{lc cc rl cc cc cc cc}
    \hline
    Star & Gaia ID & RA & Dec & $\varpi^a$ & $\sigma_{\varpi}^b$ & G & $\sigma_{G}$ & BP & $\sigma_{BP}$ & RP & $\sigma_{RP}$ & $E(B-V)$ & $[Fe/H]$ \\
     & & deg & deg & mas & mas & mag & mag & mag & mag & mag & mag & mag & dex \\
    \hline
    SW And    & 2857456207478683776 & 005.93 &  +29.40 & 1.78 & 0.16 &    &   &    &   &    &   &  0.038 & -0.24 \\
    XX And    & 0370067649378653440 & 019.36 &  +38.95 & 0.69 & 0.05 &    &   &    &   &    &   &  0.039 & -1.94 \\
    WY Ant    & 5461994297841116160 & 154.02 & -29.73 & 0.91 & 0.06 &    &   &    &   &    &   &  0.059 & -1.48\\
    X Ari     & 0015489408711727488 & 047.13 &  +10.45 & 1.84 & 0.04 &    &   &    &   &    &   &  0.180 & -2.43 \\
    AE Boo    & 1234729400256865664 & 221.90 &  +16.85 & 1.14 & 0.04 & 10.5650  & 0.0003  & 10.7366  & 0.0005  & 10.3010  & 0.0005  &  0.023 & -1.39 \\
    ST Boo    & 1374971554331800576 & 232.66 &  +35.78 & 0.74 & 0.05 & 10.9287  & 0.0003  & 11.2291  & 0.0003  & 10.6583  & 0.0003  &  0.021 & -1.76 \\
    TV Boo    & 1492230556717187456 & 214.15 &  +42.36 & 0.75 & 0.03 & 10.9110  & 0.0001  & 11.0410  & 0.0001  & 10.6846  & 0.0001  &  0.010 & -2.44 \\
    UY Boo    & 3727833391597367424 & 209.69 &  +12.95 & 0.62 & 0.05 &    &   &    &   &    &   &  0.033 & -2.56\\
    ST CVn    & 1453674738379109760 & 209.39 &  +29.86 & 0.77 & 0.03 & 11.2942  & 0.0002  & 11.4570  & 0.0002  & 11.0221  & 0.0002  &  0.012 & -1.07 \\
    UY Cam    & 1134921885080388992 & 119.75 &  +72.79 & 0.72 & 0.05 & 11.4736  & 0.0001  & 11.5893  & 0.0002  & 11.2529  & 0.0002  &  0.022 & -1.33 \\
    YZ Cap    & 6884361748289023488 & 319.88 & -15.12 & 0.85 & 0.07 & 11.2186  & 0.0001  & 11.3838  & 0.0001  & 10.9278  & 0.0001  &  0.063 & -1.06 \\
    RZ Cep    & 2211629018927324288 & 339.81 & +64.86 & 2.36 & 0.03 &  9.2562  & 0.0001  &  9.5564  & 0.0002  &  8.8052  & 0.0002  &  $0.250^c$ & -1.77 \\
    RR Cet    & 2558296724402139392 & 023.03 & +01.34 & 1.52 & 0.08 &  9.6155  & 0.0004  &  9.9355  & 0.0008  &  9.3142  & 0.0008  &  0.022 & -1.45 \\
    CU Com    & 3952883463090843520 & 186.19 &  +22.41 & 0.23 & 0.03 & 13.23737 & 0.00009 & 13.3898  & 0.0005  & 12.9105  & 0.0004  &  0.023 & -2.38  \\
    RV CrB    & 1317846466364172800 & 244.86 &  +29.71 & 0.65 & 0.03 & 11.3280  & 0.0003  & 11.4692  & 0.0002  & 11.0646  & 0.0001  &  0.039 & -1.69 \\
    W Crt     & 3546458301374134528 & 171.62 & -17.91 & 0.75 & 0.04 & 11.4505  & 0.0002  & 11.6498  & 0.0002  & 11.1385  & 0.0002  &  0.040 & -0.54 \\
    UY Cyg    & 1858568795806177792 & 314.12 &  +30.43 & 0.98 & 0.03 & 10.9762  & 0.0001  & 11.1854  & 0.0007  & 10.5660  & 0.0001  &  0.129 & -0.80 \\
    XZ Cyg    & 2142052889490819328 & 293.12 &  +56.39 & 1.57 & 0.03 &  9.6753 & 0.0002 &    &   &    &   &  0.096 & -1.44 \\
    DX Del    & 1760981190300823808 & 311.87 &  +12.46 & 1.68 & 0.03 &  9.8082  & 0.0001  & 10.0313  & 0.0008  &  9.4040  & 0.0005  &  0.092 & -0.39 \\
    SU Dra    & 1058066262817534336 & 174.49 &  +67.33 & 1.40 & 0.03 &  9.6626  & 0.0003  &  9.884   & 0.001   &  9.3647  & 0.0003  &  0.010 & -1.80 \\
    SW Dra    & 1683444631037761024 & 184.44 &  +69.51 & 1.01 & 0.03 &  10.3860 & 0.0005 &    &   &    &   &  0.014 & -1.12 \\
    CS Eri    & 4947090013255935616 & 039.27 & -42.96 & 2.07 & 0.03 &  8.9243  & 0.0001  &  9.0802  & 0.0002  &  8.6757  & 0.0002  &  0.018 & -1.41\\
    RX Eri    & 2981136563930816128 & 072.43 & -15.74 & 1.62 & 0.03 &  9.5555  & 0.0002  &  9.8422  & 0.0008  &  9.1714  & 0.0003  &  0.058 & -1.33 \\
    SV Eri    & 5165689383172441216 & 047.97 & -11.35 & 1.22 & 0.06 &    &   &    &   &    &   &  0.085 & -1.70 \\
    RR Gem    & 0886793515494085248 & 110.39 & +30.88 & 0.69 & 0.05 &    &   &    &   &    &   &  0.054 & -0.29 \\
    TW Her    & 4596935593202765184 & 268.63 & +30.41 & 0.86 & 0.02 & 11.2076  & 0.0003  & 11.5273  & 0.0003  & 10.9234  & 0.0003  &  0.042 & -0.69 \\
    VX Her    & 4467433017735910912 & 247.67 & +18.37 & 0.98 & 0.06 &  10.774 & 0.001 &    &   &    &   &  0.044 & -1.58 \\
    SV Hya    & 3499611306368945536 & 187.63 & -26.05 & 1.21 & 0.05 & 10.4409  & 0.0002  & 10.6516  & 0.0001  & 10.1201  & 0.0001  &  0.080 & -1.50 \\
    V Ind     & 6483680332235888896 & 317.87 & -45.07 & 1.50 & 0.04 &  9.8640  & 0.0002  & 10.012   & 0.001   &  9.5625  & 0.0002  &  0.043 & -1.50 \\
    BX Leo    & 3972712536822542336 & 174.51 & +16.54 & 0.59 & 0.04 & 11.5275  & 0.0003  & 11.6695  & 0.0002  & 11.2526  & 0.0004  &  0.023 & -1.28 \\
    RR Leo    &  630421935431871232 & 151.93 & +23.99 & 1.00 & 0.09 &    &   &    &   &    &   &  0.037 & -1.60 \\
    TT Lyn    & 1009665142487836032 & 135.78 & +44.59 & 1.22 & 0.04 &  9.7423  & 0.0002  &  9.847   & 0.004   &  9.3698  & 0.0006  &  0.017 & -1.56 \\
    RR Lyr    & 2125982599341232896 & 291.37 & +42.78 & -2.61 & 0.61 &    &   &    &   &    &   &  0.030 & -1.39 \\
    RV Oct    & 5769986338215537280 & 206.63 & -84.40 & 1.01 & 0.03 & 10.8412  & 0.0001  & 11.1041  & 0.0002  & 10.3840  & 0.0002  &  0.180 & -1.71 \\
    UV Oct    & 5768557209320424320 & 248.10 & -83.90 & 1.89 & 0.03 &  9.2246  & 0.0006  &  9.7825  & 0.0007  &  9.0224  & 0.0008  &  0.091 & -1.74 \\
    AV Peg    & 1793460115244988800 & 328.01 & +22.57 & 1.46 & 0.03 & 10.4721  & 0.0002  & 10.6924  & 0.0003  & 10.0693  & 0.0002  &  0.067 & -0.08 \\
    BH Peg    & 2828497068363486720 & 343.25 & +15.79 & 1.11 & 0.04 &    &   &    &   &    &   &  0.077 & -1.22 \\
    DH Peg    & 2720896455287475584 & 333.86 & +06.82 & 2.07 & 0.05 &    &   &    &   &    &   &  0.080 & -0.92 \\
    RU Psc    & 0294072906063827072 & 018.61 & +24.42 & 0.91 & 0.08 &    &   &    &   &    &   &  0.043 & -1.75 \\
    BB Pup    & 5707380936404638336 & 126.09 & -19.54 & 0.59 & 0.04 & 12.0508  & 0.0001  & 12.2939  & 0.0003  & 11.6624  & 0.0002  &  0.105 & -0.60 \\
    HK Pup    & 3030561879348972544 & 116.20 & -13.10 & 0.70 & 0.04 & 11.1836  & 0.0002  & 11.4345  & 0.0003  & 10.7628  & 0.0004  &  0.160 & -1.11 \\
    RU Scl    & 2336550174250087936 & 000.70 & -24.95 & 1.08 & 0.07 & 10.1333  & 0.0009  & 10.457   & 0.002   &  9.9199  & 0.0006  &  0.018 & -1.27 \\
    SV Scl    & 5022411786734718208 & 026.25 & -30.06 & 0.50 & 0.04 & 11.3036  & 0.0002  & 11.4445  & 0.0001  & 11.0594  & 0.0001  &  0.014 & -1.77 \\
    AN Ser    & 1191510003353849472 & 238.38 & +12.96 & 0.95 & 0.04 & 10.8466  & 0.0002  & 11.0508  & 0.0004  & 10.4944  & 0.0002  &  0.040 & -0.07 \\
    AP Ser    & 1167409941124817664 & 228.50 & +09.98 & 0.75 & 0.04 &    &   &    &   &    &   &  0.042 & -1.58 \\
    T Sex     & 3846786226007324160 & 148.37 & +02.06 & 1.25 & 0.04 & 9.9636 & 0.0008 &    &   &    &   &  0.044 & -1.34 \\
    V0440 Sgr & 6771307454464848768 & 293.09 & -23.85 & 1.40 & 0.04 & 10.1576  & 0.0006  & 10.638   & 0.001   &  9.922   & 0.002   &  0.085 & -1.40 \\
    V0675 Sgr & 4039386574037718528 & 273.40 & -34.32 & 1.11 & 0.06 & 10.1923  & 0.0002  & 10.5190  & 0.0005  &  9.8221  & 0.0003  &  0.130 & -2.28 \\
    MT Tel    & 6662886605712648832 & 285.55 & -46.65 & 1.93 & 0.04 &  8.9110  & 0.0009  &  9.096   & 0.002   &  8.674   & 0.001   &  0.038 & -1.85 \\
    AM Tuc    & 4692528057537147136 & 019.63 & -67.92 & 0.54 & 0.03 & 11.5381  & 0.0002  & 11.7086  & 0.0002  & 11.2378  & 0.0003  &  0.023 & -1.49 \\
    AB UMa    & 1546016672688675200 & 182.81 & +47.83 & 0.98 & 0.03 & 10.78757 & 0.00008 & 11.05295 & 0.00007 & 10.3952  & 0.0001  &  0.022 & -0.49 \\
    RV UMa    & 1561928427003019520 & 203.33 & +53.99 & 0.92 & 0.03 & 10.6933  & 0.0002  & 10.8908  & 0.0002  & 10.4122  & 0.0002  &  0.018 & -1.20 \\
    SX UMa    & 1565435491136901888 & 201.56 & +56.26 & 0.75 & 0.04 & 10.7786  & 0.0002  & 10.9016  & 0.0002  & 10.5589  & 0.0002  &  0.010 & -1.81 \\
    TU UMa    & 4022618712476736896 & 172.45 & +30.07 & 1.56 & 0.06 &    &   &    &   &    &   &  0.022 & -1.51 \\
    UU Vir    & 3698725337375614464 & 182.15 & -00.46 & 1.21 & 0.08 & 10.5155 & 0.0002 &    &   &    &   &  0.018 & -0.87 \\
    \hline
    \multicolumn{13}{l}{$^a$ From \emph{Gaia} DR2 archive; without 0.03 mas offset.}\\
    \multicolumn{13}{l}{$^b$ Formal uncertainty from \emph{Gaia} DR2 archive. We adopt $1.08\sigma_{\varpi}$ in this work.}\\
    \multicolumn{13}{l}{$^c$ From \protect\citet{fernley_absolute_1998}.}
    \end{tabular}
\end{table*}

\section{The Calibrators} \label{sec:obs}

We adopted the sample of 55 Galactic RRL observed as part of CRRP, that cover the characteristic range of period and metallicity and are well distributed on the sky, as the calibrators of the PL, LZ, and PLZ relations. A detailed summary of the photometry and demographics of the pulsation properties of this sample was presented in Paper I, and we refer the reader to this work, particularly their Tables 1 and 5 for the periods and mean magnitudes of the stars in our sample. The metallicities adopted are originally from \protect\citet{fernley_absolute_1998}, and have typical uncertainties of $\pm0.15$ dex. They are assumed to be on a scale comparable to that of \protect\citet{zinn_globular_1984}, since no significant offset is found for the stars in common with \protect\citet{Layden_metallicities_1994}. Empirical light curves are available in Paper I for the Johnson $UBV$, Kron-Cousins $RI$, 2MASS $JHK_s$, and \emph{Spitzer} $[3.6]$ and $[4.5]$ passbands, and from the \emph{Gaia} archive for $G, BP, RP$. For stars with insufficient light curve coverage (a significant fraction of our sample in the $URJHK_s$ bands), we have used average magnitudes derived from template light curves, which will be presented in a forthcoming work (Beaton et al. in prep). In this work, two stars were removed from the calibrator sample. RR Lyr, the nearest and brightest RRL, had a $G$ magnitude measurement that was clearly corrupted, leading to the star's large negative parallax in DR2 \protect\citep{gaia_collaboration_gaia_2018-1}. The second star removed was CU Com, the only double mode RRL in our sample. Furthermore, at $\sim 4$kpc, it is twice as distant than any other star in our sample and therefore its distance modulus is proportionally more affected by uncertainties in the \emph{Gaia} zero-point offset given its small parallax. 

\subsection{Extinction}
For most of the stars in our sample, reddening values are available from \protect\citet{feast_luminosities_2008}, which were estimated using a 3D Galactic extinction model. For the two stars (BB Pup and DH Peg) without reddenings, we adopted the value from the reddening map in \protect\citet{schlafly_measuring_2011}.
Uncertainties of the individual reddening values, however, were not available, and we adopt $\sigma_{E(B-V)} = 0.16 E(B-V)$, as suggested by \protect\citet{schlegel_maps_1998}. The reddenings from \protect\citet{feast_luminosities_2008} were converted to extinction by assuming $A_V = 3.1E(B-V)$ and adopting the extinction ratios defined by \protect\citet{cardelli_relationship_1989}, and extended in the the mid-infrared according to \protect\citet{indebetouw_wavelength_2005}. Table~\ref{tab:ext} lists our adopted extinction ratios, as well as the assumed effective wavelength for each of our filters. 

\begin{table*}
    \centering
    \label{tab:ext}
    \caption{Extinction Ratios}
    \begin{tabular}{l ccccc  ccc ccc cc}
    \hline
     & U & B & BP & V & R & G & RP & I & J & H & K & [3.6] & [4.5] \\
    \hline
    $\lambda [\mu m]^a$ & 0.366 & 0.436 & 0.532 & 0.545 & 0.641 & 0.673 & 0.797 & 0.798 & 1.235 & 1.662 & 2.159 & 3.545 & 4.442 \\
    $A_{\lambda}/E(B-V)$ & 4.821 & 4.151 & 3.221 & 3.128 & 2.607 & 2.455 & 1.866 & 1.860 & 0.893 & 0.552 & 0.363 & 0.208 & 0.174 \\
    \hline
    \multicolumn{14}{p{.9\linewidth}}{$^a$ Central wavelengths from \protect\citet{bessell_standard_2005} ($UBVRI$), \protect\citet{jordi_gaia_2010} (Gaia $G,BP,RP$), \protect\citet{cohen_spectral_2003}, (2MASS $JHK_s$), and \protect\citet{indebetouw_wavelength_2005} (IRAC $[3.6],[4.5]$)}
    \end{tabular}
\end{table*}

\begin{figure}
\includegraphics[scale=0.73]{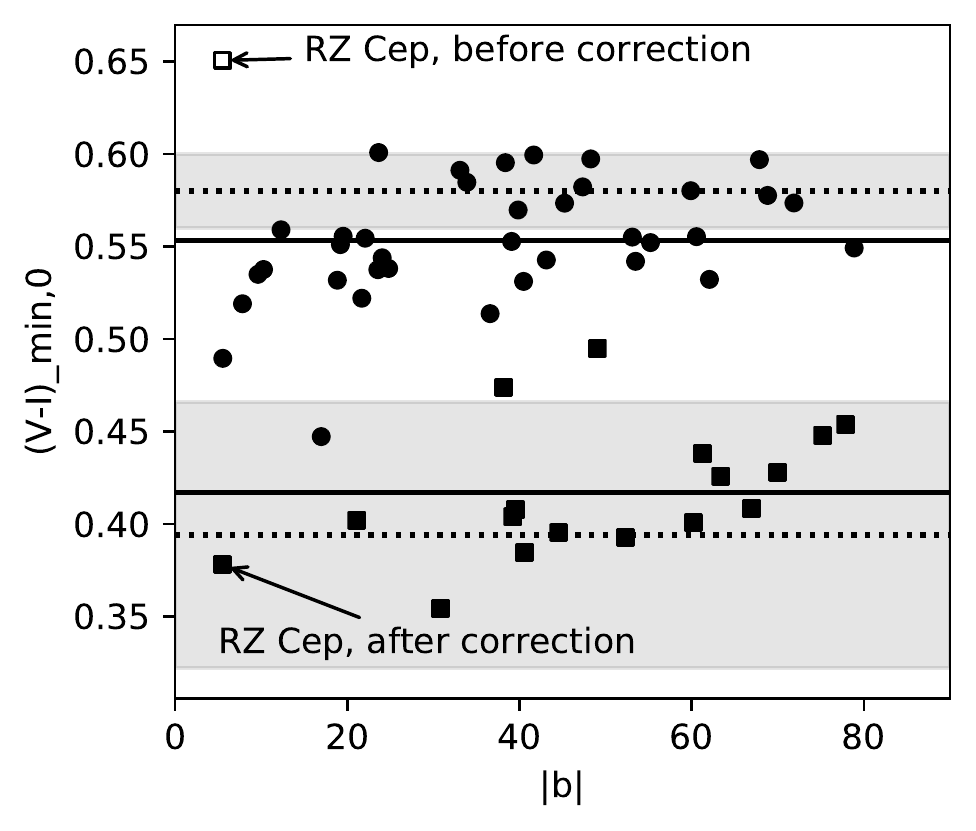}
\centering
\caption{Reddening corrected minimum light color versus Galactic latitude for all stars in our sample. RRab and RRc stars are represented with circles and squares, respectively. The dotted lines are the canonical mean values and the shaded region is their uncertainty. The solid line is the mean value of our sample. The open square shows the minimum light color for RZ Cep using the original underestimated extinction.}
\label{fig:red} 
\end{figure}

To check the accuracy of these reddenings for individual stars, we determined the $(V-I)$ color at minimum light, which is an established reddening indicator for RRab stars \protect\citep{sturch_intrinsic_1966, mateo_optical_1995, guldenschuh_intrinsic_2005}, and potentially for RRc stars \protect\citep{layden_colors_2013}. Since we do not have simultaneous photometry in the $V$ and $I$ bands to compute the minimum light color, we elected to use our smoothed light curves (see Paper I for details on the how the smoothed light curves were obtained) to compute a color curve for each star and applied the extinction correction. For RRab stars, the quantity $(V-I)_{0,min}$ is defined as the mean extinction-corrected color in the phase interval of $0.5 - 0.8$ \protect\citep{sturch_intrinsic_1966}. As shown in Figure~\ref{fig:red}, our average value of $(V-I)_{0,min} = 0.55\pm0.03$ mag for the RRab stars is in agreement with the canonical value \protect\citep[$0.58\pm0.02$,][]{guldenschuh_intrinsic_2005}, and since we detect no obvious trends with $[Fe/H]$ or Galactic latitude, we determine that the \protect\citet{feast_luminosities_2008} reddening values are appropriate. For the RRc stars, $(V-I)_{0,min}$ is defined in the phase interval of $0.45 - 0.65$ \protect\citep{layden_colors_2013}. In both $(V-I)_{0,min}$ versus period and Galactic latitude, only RZ Cep is an outlier, with an extremely underestimated extinction. Therefore, we adopt the higher value of $E(B-V)=0.25$ for this star from \protect\citet{fernley_absolute_1998}, which results in a $(V-I)_{min,0}$ equal to the mean values of the other RRc stars. The final $V$ band extinction values used in this work are given in Table~\ref{tab:stars}.

\begin{figure}
\includegraphics[scale=0.85]{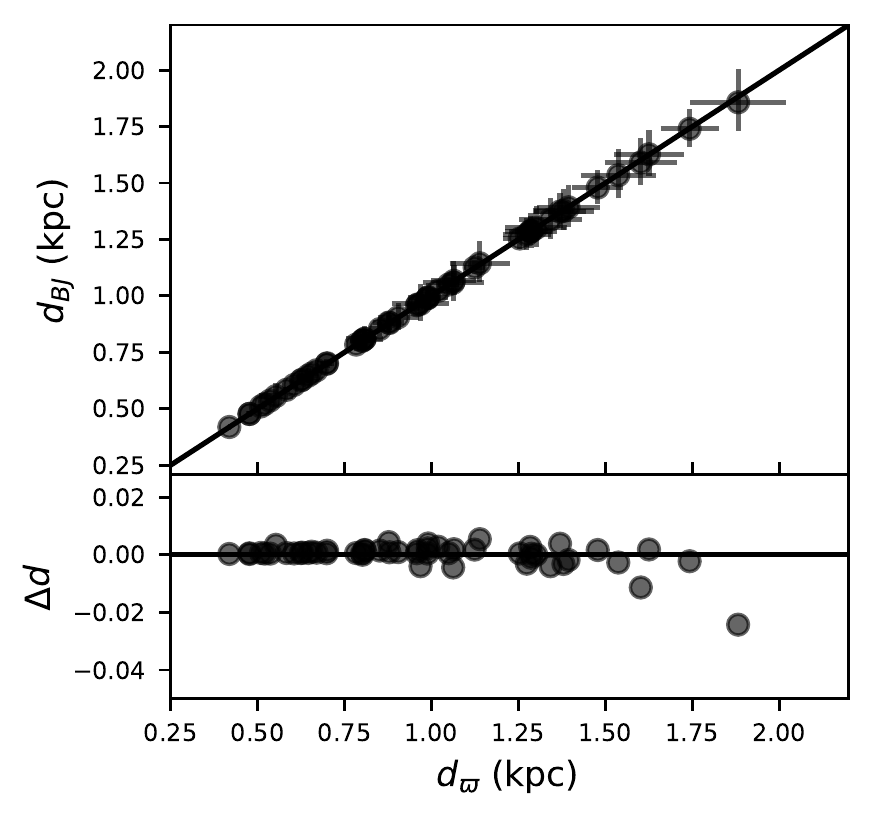}
\centering
\caption{Comparison between the distance derived via inversion of parallax to Bayesian-inferred distances from \protect\citet{bailer-jones_estimating_2018}. The residuals are shown in the bottom panel. The same global zero-point offset used in the Bayesian method, 0.03 mas, was applied before inversion. For all the stars in our sample, inversion of parallax produces distances that are indistinguishable from the Bayesian method. }
\label{fig:pi2} 
\end{figure}

\subsection{Gaia DR2}\label{sec:dr2}
We identified the closest match in RA/Dec in the \emph{Gaia} DR2 source catalog and recovered parallaxes for each of our stars. To ensure that we positively identified our stars in DR2, we also downloaded the time series data\footnote{Accessed from \href{http://gaia.ari.uni-heidelberg.de/singlesource.html}{the ARI website.}} and phase-folded the \emph{Gaia} light curve with the periods derived from our own data (Paper I). Time series data was not available for two stars (AP Ser and TU UMa), and for these stars we compared the average $G$ and $V$ magnitudes to confirm a positive match. An investigation of the light curves shows that our periods did not properly phase the \emph{Gaia} data for two additional stars (BX Leo and RU Psc), but the coverage in DR2 is too sparse to draw any strong conclusions. Since these stars are not outliers in the PL relations, we assumed our periods (which were computed from many more epochs) are correct. The \emph{Gaia} source IDs, parallaxes, and intensity averaged mean magnitudes (from the variable catalog where available) for the stars in our sample are given in Table~\ref{tab:stars}. The formal parallax uncertainties in the \emph{Gaia} catalog reflect only the internal consistency of the measurement, and could be underestimated by as much as 30\% for the magnitude range of our sample \protect\citep{lindegren_gaia_2018}. The \emph{Gaia} archive also provides an astrometry quality parameter, RUWE, to help identify spurious parallaxes\footnote{See the presentation available on the \emph{Gaia} known issues page for more details \href{https://www.cosmos.esa.int/web/gaia/dr2-known-issues\#AstrometryConsiderations}{here}}. All of the sources in our sample have an RUWE < 1.4, indicating there are no obvious issues with their astrometry.

The parallaxes in DR2 are known to suffer from a global zero-point offset \protect\citep{lindegren_gaia_2018, arenou_gaia_2018}. Furthermore, the derived offset varies when tested using various populations of stars, and is likely a function of sky position, magnitude, and color. A systematic offset of $\Delta \varpi = -0.03$ mas (where $\Delta \varpi = \varpi_{Gaia} - \varpi_{true}$) is measured using quasi-stellar objects (QSOs), but it can range from $-0.030$ to $-0.056$ mas for RRL \protect\citep[see Table 1 from][]{arenou_gaia_2018}. Given the severely limited sample of RRL with previous parallax estimates, it is not possible to derive an independent zero-point for RRL (i.e. without assuming a PL relation). Therefore, in this work we provisionally elect to adopt the global offset of -0.03 mas for consistency with other current works, and discuss the impact of this decision further in Section~\ref{sec:zp}. 

It has been shown that deriving distances from a simple inversion of the parallax may result in large biases \protect\citep{gaia_collaboration_gaia_2017, luri_gaia_2018}. The stars in our sample, however, are all relatively nearby, and so we are generally dealing with low fractional uncertainty in the parallax ($\sigma_{\varpi}/\varpi < 10\%$). In Fig.~\ref{fig:pi2}, we compare distances of our stars derived via inversion of parallax to the Bayesian method from \protect\citet{bailer-jones_estimating_2018}. We note that in \protect\citet{bailer-jones_estimating_2018}, the authors applied the same global zero-point offset, 0.03 mas, that is adopted for this work. Clearly, within 2 kpc, inversion produces results identical to the Bayesian method, provided the same global zero-point offset has been applied in both cases. In what follows, we use the direct and prior-independent method of inversion to calculate the absolute magnitudes of stars in this work.

\section{Results}
Theoretical investigations into the RRL PL (PW) have long predicted a dependence on metallicity in all passbands, generally with slopes on the order of 0.2 mag/dex \protect\citep{bono_pulsational_2003, marconi_new_2015, neeley_new_2017}. Prior to \emph{Gaia} DR1 however, empirical studies found either metallicity slopes that are significantly smaller \protect\citep[e.g.][]{dambis_mid-infrared_2014} or not measurable \protect\citep[e.g.][]{madore_preliminary_2013}. This is due in part to the difficulty in measuring $[Fe/H]$ of RRL, and in part due to the limited number of RRL with independent distance measurements prior to \emph{Gaia}. More recently, stronger metallicity slopes have been found in studies utilizing \emph{Gaia} parallaxes, for example $0.17\pm0.10$ mag dex$^{-1}$ in \protect\citet{sesar_probabilistic_2017} and $0.17\pm0.03$ mag dex$^{-1}$ in \protect\citet{muraveva_rr_2018}.

In the following sections, we compute PW and PWZ relations for two and three band combinations of our 13 passbands. We also compute PL and PLZ relations for the $I$ band and longer. While we do find measurable PL relations in the shorter bands, these are much more likely to be sample dependent, and we recommend utilizing the PW(Z) relations when using optical wavelengths. The relations were determined using either bivariate or trivariate weighted least squares fits. Absolute magnitudes were determined using the equation $M_{\lambda} = m_{\lambda} - A_{\lambda} -10 +5\log (\varpi)$, where $\varpi$ is the \emph{Gaia} parallax in mas. The uncertainty in the absolute magnitude includes the uncertainties in photometry, parallax, and extinction combined following standard error propagation, and the weights in the fit are $1/\sigma^2$. In all bands, the parallax is the dominant source of uncertainty in the absolute magnitudes. Period and metallicity uncertainties were not included in the weights, since the period uncertainties are negligible and uncertainties in $[Fe/H]$ are assumed to be the same for all stars and therefore equal in weight. The periods of the first overtone (RRc) stars have been transformed to their equivalent fundamental period, or ``fundamentalized'', according to the relation $\log P_F = \log P_{FO} + 0.127$ \protect\citep{iben_comments_1971, rood_metal-poor_1973, cox_double-mode_1983}. Weighted least squares fits were performed on the combined RRab and fundamentalized RRc sample. To decouple the error on the zero-point and slopes, we fit around the mean period and metallicity, so our PW and PL relations take the form
\begin{eqnarray}
M = a + b (\log P_F + 0.30)
\end{eqnarray}
and the PWZ and PLZ relations take the form
\begin{eqnarray}
M = a + b(\log P_F + 0.30) + c([Fe/H] + 1.36).
\end{eqnarray}

We elect to perform a jackknife resampling test due to the heterogeneous nature of our absolute magnitude uncertainties. The total uncertainty is almost always dominated by the parallax uncertainty such that weighting will tend to stress the stars with better parallaxes. The jackknife test is performed by removing one star from our sample and repeating the measurement. As such, it tests for how likely our measurement is to be impacted by a small number of well measured points.

We also explored two additional fitting techniques, a robust analysis and a Bayesian analysis (see the Appendix for details). The robust fitting technique method reweights the data points according to the scatter on the fit, in order to mitigate the influence of outlying data points. Therefore, this method does not take into account the strong range of uncertainties in our data set. As a result, we see steeper period slopes, albeit such changes are largely modulated by a change in the zero-point, such that the end results are similar. The Bayesian technique allows us to explore the influence of non-negligible uncertainties in the metallicity, which are not accounted for in a traditional linear regression. While the Bayesian method does predict slightly steeper metallicity slopes, we do not find a significant difference in the final results. Thus, we opt for our more conservative approach of a traditional regression that weights the data points relative to their uncertainty, but provide the results of the alternative fits in the Appendix.

\begin{figure*}
\includegraphics[scale=0.8]{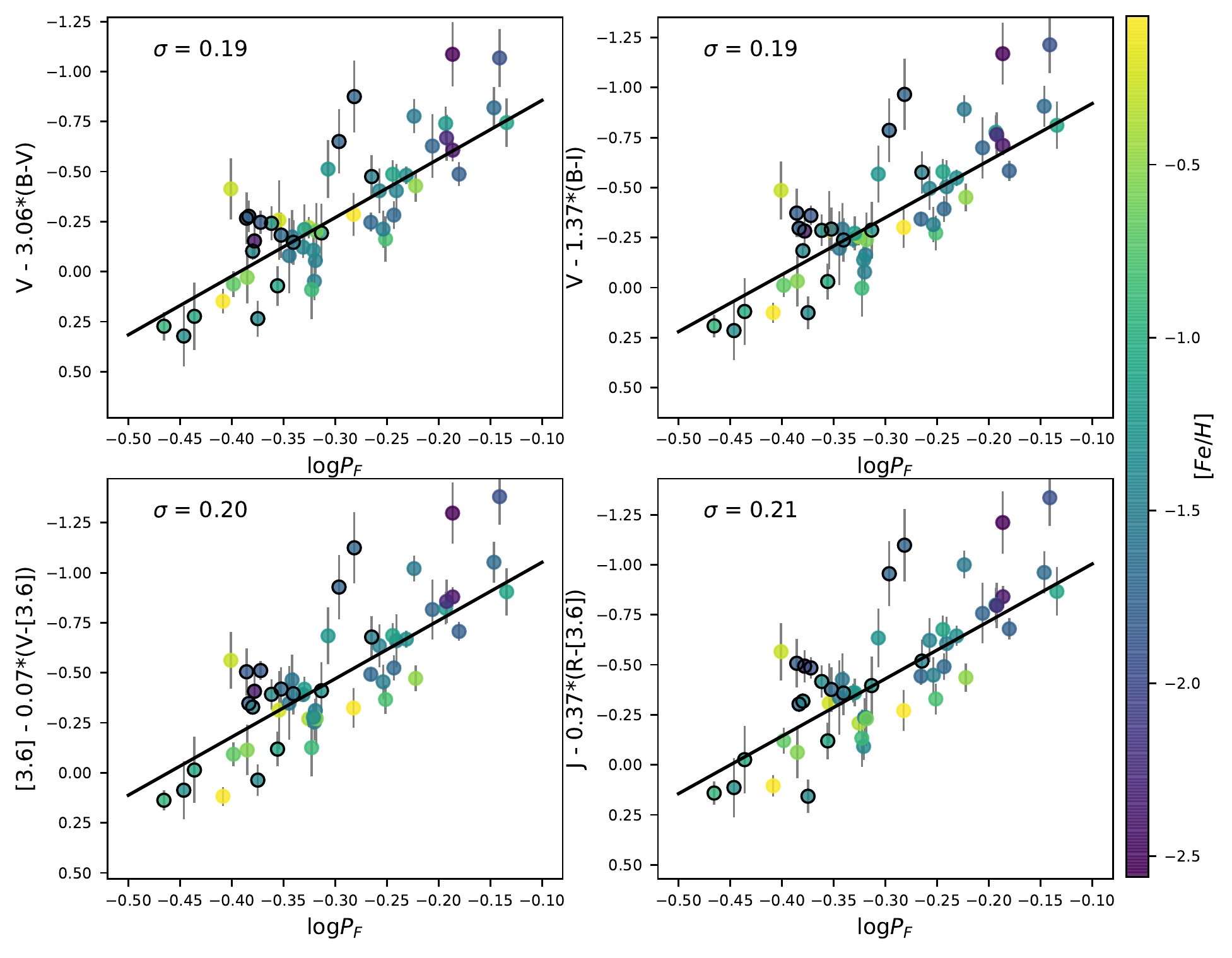}
\centering
\caption{Period-Wesenheit relations for two-band (left panels) and three-band (right panels) combinations. The periods of RRc stars have been fundamentalized, and are outlined in black circles. The points have also been color coded according to metallicity, and there is a visible trend in the residuals with metallicity. The residuals are also highly correlated in all filter combinations, with is consistent with the scatter dominated by parallax uncertainties. The sigma quoted in the top right corner of each panel is the RMS about the fitted relation.}
\label{fig:pw} 
\end{figure*}

\begin{figure*}
\includegraphics[scale=0.80]{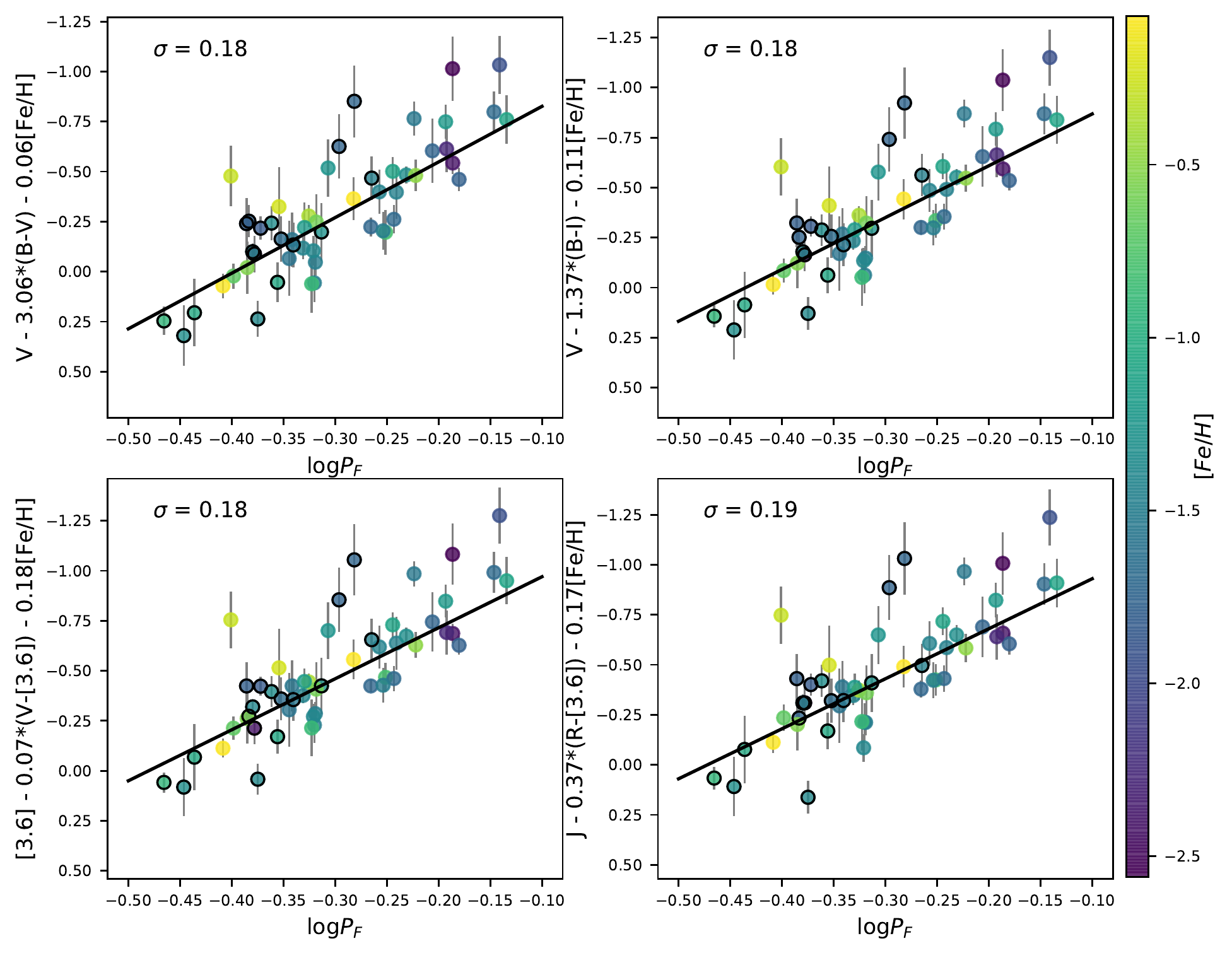}
\centering
\caption{Same as Fig.~\ref{fig:pw}, but now for Period-Wesenheit-Metallicity relations. The metallicity term has been subtracted from the Wesenheit magnitude in order to demonstrate the reduced scatter. The residuals are still highly correlated between different band combinations, indicating that the scatter is dominated by distance uncertainties.}
\label{fig:pwz} 
\end{figure*}

\subsection{Wesenheit magnitude relations}\label{sec:pwz}

Period-Wesenheit relations are designed to provide distance estimates for individual stars, without knowledge of their specific extinctions. The Wesenheit magnitude \protect\citep{madore_period-luminosity_1982} is defined by 
\begin{eqnarray}
W(M_1, M_2-M_3) = M_1 - \alpha (M_2 - M_3)
\end{eqnarray}
where the color coefficient $\alpha$ is determined directly from the reddening law ($\alpha = A_1/(A_2-A_3)$). We adopt the reddening law from \protect\citep{cardelli_relationship_1989}, with $R_V = 3.1$, but note that infrared PW relations are only weakly sensitive to variations in the dust parameter \protect\citep{subramanian_young_2017}. For the two-band combinations, $M_1 = M_3$, and is the redder of the two bands. For three band combinations, $M_2$ and $M_3$ are the bluest and reddest bands, respectively.  

Coefficients for a subset of the possible two and three band PW relations are given in Table~\ref{tab:results}. In this Table, we do not include any filter combinations that include the \emph{Gaia} bands, because they are derived from a smaller sample of stars. The full version of Table~\ref{tab:results} is available in the online edition, and does include combinations using the \emph{Gaia} bands. In Fig.~\ref{fig:pw}, we show four of the PW relations; two purely optical combinations (top panels) and two combinations of optical and infrared bands (bottom panels). The points are color coded according to their metallicity. The most striking feature of these plots is that the residuals are practically identical for all filter combinations. This is consistent with the scatter being dominated by parallax uncertainties, which we discuss further in Section \ref{sec:scatter}. 

The residuals from the PW relations are also correlated with metallicity, with the more metal-poor stars being systematically brighter, suggesting a PWZ relation would better model our data. The coefficients of two and three band PWZ relations are listed in Table~\ref{tab:results}. The coefficient of the metallicity term ranges from 0.06 for $W(V,B-V)$ to 0.2 for $W([4.5],[3.6]-[4.5])$, and agree well within $1\sigma$ with values derived from pulsation models \protect\citep{marconi_new_2015}. Fig.~\ref{fig:pwz} shows the PWZ relations, with the metallicity term subtracted from the y-axis, in the same band combinations as Fig.~\ref{fig:pw}. In all cases, the PWZ relations do have slightly smaller dispersion, but remain limited by distance uncertainties. 

Comparisons of PW and PWZ relations with previous studies can be complicated given their dependence on the adopted reddening law. Even authors using the same reddening law can infer different color coefficients for the Wesenheit magnitude, if they assume different central wavelengths of the same filter. For instance \protect\citet{braga_distance_2015} measured PW relations in the globular cluster M4, but we cannot directly compare our slopes since their relations are tied to a reddening law specific to M4. The slopes and zero-points of our PWZ relations are however entirely consistent with theoretical models \protect\citep{marconi_new_2015, neeley_new_2017}, when we correct for slight differences in the adopted $\alpha$ parameter; only combinations involving the $U$ band disagree by more than $2\sigma$.

\subsection{Standard magnitude relations}\label{sec:plz}

Historically, PL(Z) relations have only been considered at infrared wavelengths, while luminosity-metallicity relations have been employed in the optical (particularly the $V$ band). The reasons for this are demonstrated in the theoretical investigation of \protect\citet{catelan_rr_2004}. In the optical bands, the mixture of ZAHB and evolved RRL results in very large scatter in the period-magnitude plane, while in the infrared, the ZAHB and evolved stars all fall on the same, tight relation. From an empirical standpoint, the implications are that the slope measured in optical bands is highly dependent on the evolutionary status of the stars in your sample. Because our sample is composed of field stars, we have no way of determining their evolutionary status. As a result, we provide PL and PLZ relations only for the $I$ and longer bands, consistent with \protect\citet{catelan_rr_2004}. 

\begin{figure*}
\includegraphics[scale=0.85]{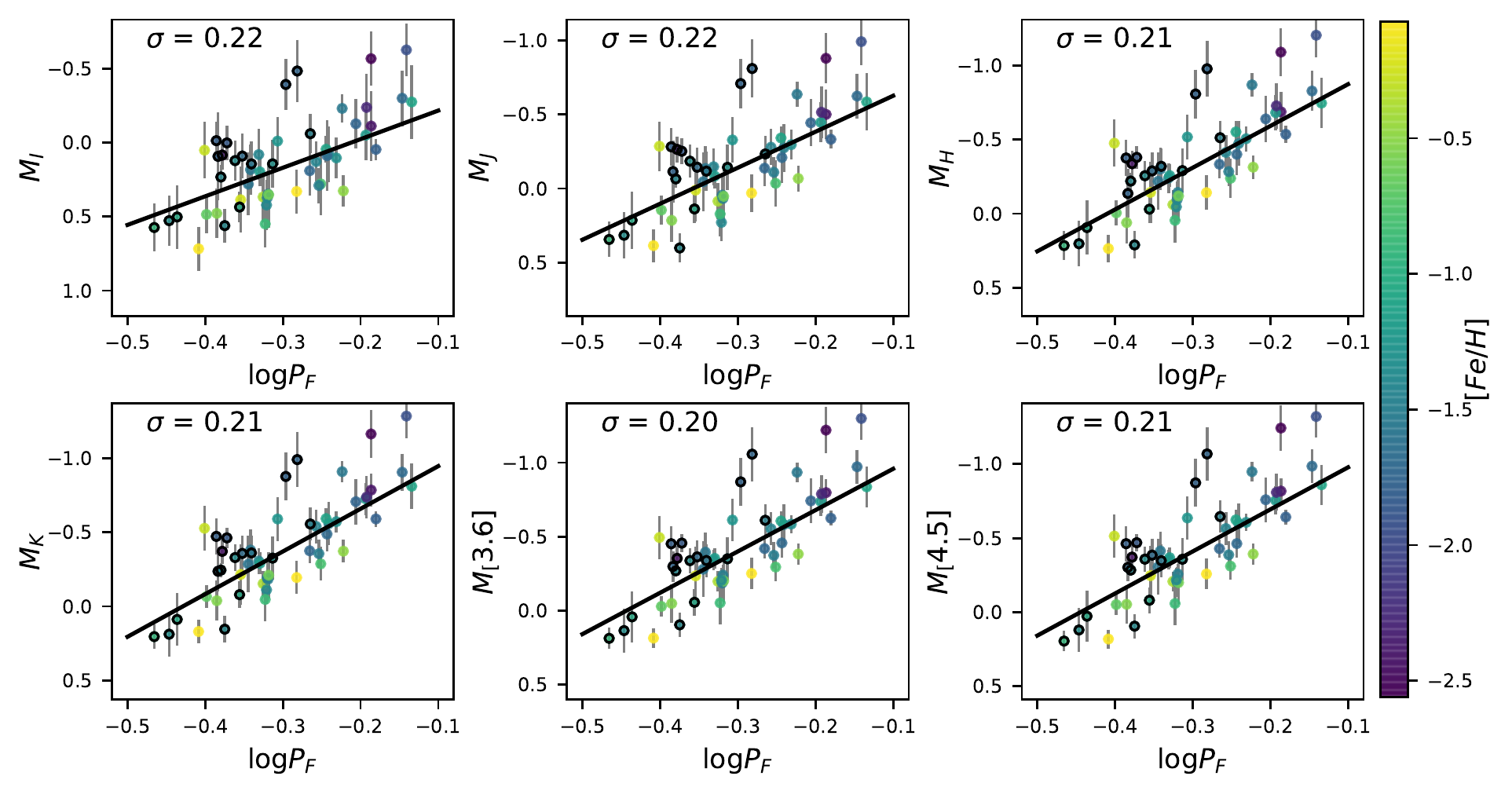}
\centering
\caption{Period-luminosity relations (PL) for the $I$ and longer passbands. Points are color coded by metallicity and RRc stars are outlied in black. The slope becomes slightly steeper moving from the $I$ to $H$ band. As with the PW relations, you can see the trend in the residuals with metallicity, where more metal rich stars are systematically fainter than the PL relation. }
\label{fig:pl} 
\end{figure*}

\begin{figure*}
\includegraphics[scale=0.85]{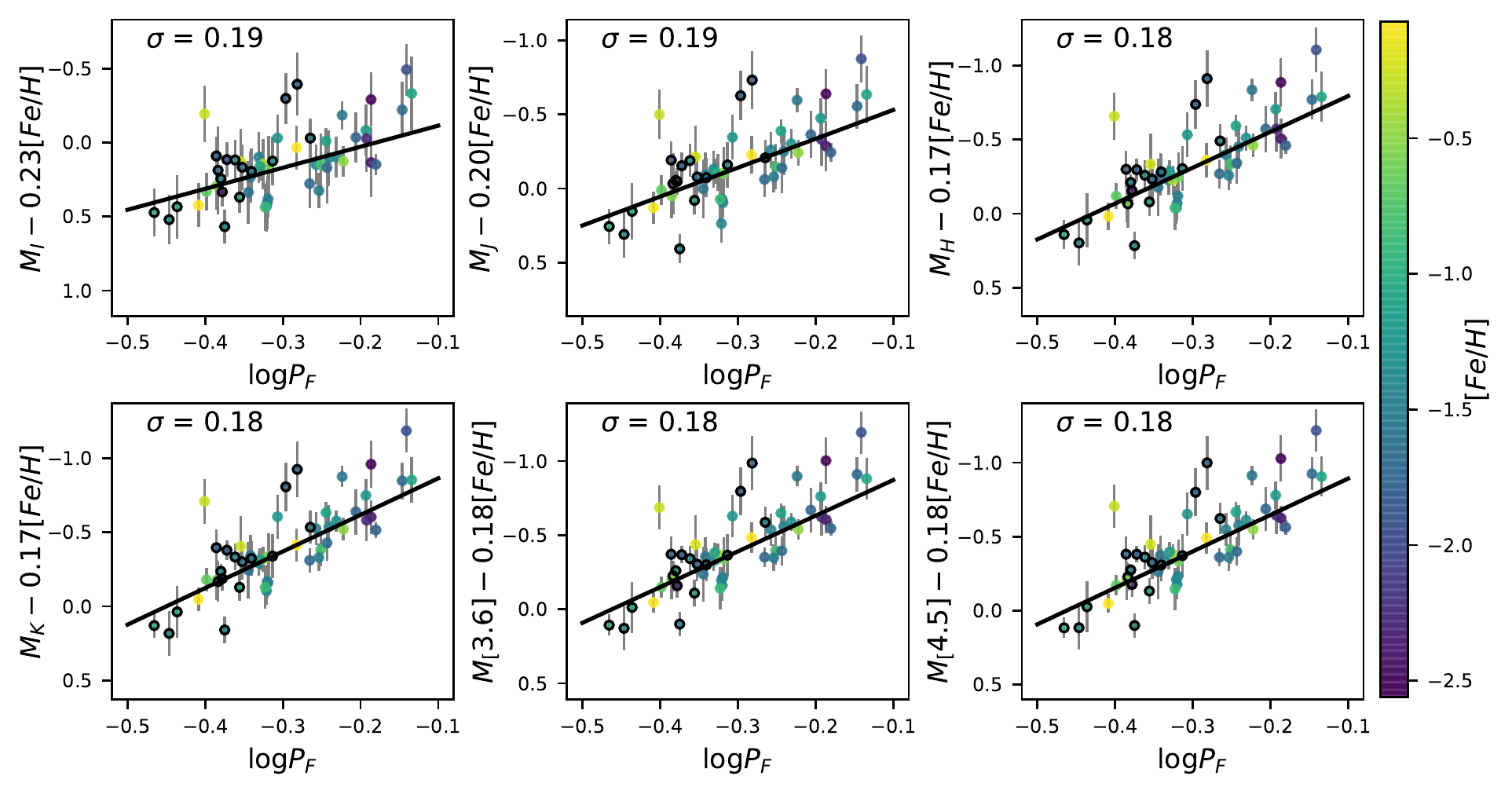}
\centering
\caption{Same as Fig.~\ref{fig:pl}, but now for the period-luminosity-metallicity relations (PLZ). The metallicity term has been subtracted from the absolute magnitudes as indicated on the y-axis. The observed scatter in the PLZ fits is slightly smaller than the PL relations, but the residuals in each band are highly correlated. }
\label{fig:plz} 
\end{figure*}

 In Fig.~\ref{fig:pl}, we show the PL relation in each band, with the points color-coded according to their metallicity. As with the PW relations, the residuals are equivalent at all wavelengths, and there is a clear trend with metallicity. Fig.~\ref{fig:plz} shows the results of the PLZ fit, and the metallicity term has been subtracted from the vertical axis. The coefficients of the PL and PLZ relations are provided in Table \ref{tab:results}. The dispersion from the PLZ relation is slightly reduced at all wavelengths when compared to the PL relation (e.g. 0.22 vs 0.19 mag in $I$ and 0.21 vs 0.18 mag in $[4.5]$).

 Overall, our PL(Z) relations are consistent with previous studies, both empirical and theoretical. We find systematically steeper slopes than recent theoretical models \protect\citet{catelan_rr_2004, marconi_new_2015, neeley_new_2017}, but the slopes are consistent within $1\sigma$. One interesting difference is that theoretical studies predict that the metallicity slope to be almost wavelength-independent, while empirical results show a steepening of the metallicity dependence at shorter wavelengths. We note, however, that the mean magnitudes derived from theoretical models are more unreliable at optical wavelengths, as the increased number of spectral features makes transforming the bolometric magnitude into the observational plane more uncertain.   
 
 On the empirical side, \protect\citet{muraveva_rr_2018} measured the coefficients of the PLZ relation in the $K$ and \emph{WISE} $W1$ bands using \emph{Gaia} DR2 data. Their sample of almost 400 RRL is based on the work of \protect\citet{dambis_rr_2013}, who compiled photometry, periods, metallicities, and extinctions from the literature. Although they adopt a larger \emph{Gaia} zero-point offset than we do (0.056 mas versus 0.03 mas), our $K$ and $[3.6]$ PLZ relations agree within the $1\sigma$ uncertainties (see their Table 4). 
 
\begin{table}
    \scriptsize
    \centering
    \caption{Weighted Least Squares Fitting Results}
    \label{tab:results}
    \begin{tabular}{l c cccc}
    \hline 
    Filters & $\alpha^a$ & a & b & c & $\sigma$ \\
    \hline
    \multicolumn{6}{c}{\emph{PL Relations$^b$}} \\
    $I     $ & &  $ 0.17\pm0.03$ & $-1.92\pm0.41$ & &  $0.22$ \\
    $J     $ & & $-0.14\pm0.03$ & $-2.38\pm0.38$ & & $0.22$ \\
    $H     $ & & $-0.31\pm0.03$ & $-2.79\pm0.34$ & & $0.21$ \\
    $K     $ & & $-0.37\pm0.03$ & $-2.84\pm0.35$ & & $0.21$ \\
    $[3.6] $ & & $-0.40\pm0.03$ & $-2.78\pm0.38$ & & $0.21$ \\
    $[4.5] $ & & $-0.41\pm0.03$ & $-2.83\pm0.39$ & & $0.21$ \\
    \multicolumn{6}{c}{\emph{PLZ Relations$^c$}} \\
    $I     $ & &  $ 0.17\pm0.03$ & $-1.40\pm0.30$ & $ 0.23\pm0.04$ & $0.19$ \\
    $J     $ & & $-0.14\pm0.02$ & $-1.91\pm0.29$ & $ 0.20\pm0.03$ & $0.19$ \\
    $H     $ & & $-0.31\pm0.02$ & $-2.40\pm0.29$ & $ 0.17\pm0.03$ & $0.18$ \\
    $K     $ & & $-0.37\pm0.02$ & $-2.45\pm0.28$ & $ 0.17\pm0.03$ & $0.18$ \\
    $[3.6] $ & & $-0.39\pm0.02$ & $-2.40\pm0.27$ & $ 0.18\pm0.03$ & $0.18$ \\
    $[4.5] $ & & $-0.40\pm0.02$ & $-2.45\pm0.28$ & $ 0.18\pm0.03$ & $0.18$ \\
    \multicolumn{6}{c}{\emph{Two-band PW Relations$^b$}} \\
    $B,U-B         $ & $ 6.228$ & $ 0.30\pm0.05$ & $ 0.94\pm0.58$ & & $0.40$ \\
    $V,B-V         $ & $ 3.058$ & $-0.27\pm0.02$ & $-2.93\pm0.25$ & & $0.19$ \\
    $R,B-R         $ & $ 1.689$ & $-0.41\pm0.02$ & $-3.21\pm0.30$ & & $0.19$ \\
    $I,V-I         $ & $ 1.467$ & $-0.42\pm0.02$ & $-2.92\pm0.30$ & & $0.20$ \\
    $J,V-J         $ & $ 0.399$ & $-0.43\pm0.03$ & $-2.88\pm0.34$ & & $0.22$ \\
    $H,J-H         $ & $ 1.618$ & $-0.59\pm0.03$ & $-3.61\pm0.30$ & & $0.20$ \\
    $[3.6],V-[3.6] $ & $ 0.071$ & $-0.47\pm0.03$ & $-2.91\pm0.37$ & & $0.20$ \\
    $[4.5],K-[4.5] $ & $ 0.918$ & $-0.44\pm0.03$ & $-3.04\pm0.40$ & & $0.21$ \\
    \multicolumn{6}{c}{\emph{Two-band PWZ Relations$^c$}} \\
    $B,U-B         $ & $ 6.228$ & $ 0.30\pm0.05$ & $ 1.71\pm0.68$ & $ 0.20\pm0.10$ & $0.37$ \\
    $V,B-V         $ & $ 3.058$ & $-0.27\pm0.02$ & $-2.78\pm0.27$ & $ 0.06\pm0.03$ & $0.18$ \\
    $R,B-R         $ & $ 1.689$ & $-0.41\pm0.02$ & $-2.99\pm0.30$ & $ 0.09\pm0.04$ & $0.18$ \\
    $I,V-I         $ & $ 1.467$ & $-0.41\pm0.02$ & $-2.60\pm0.25$ & $ 0.13\pm0.03$ & $0.18$ \\
    $J,V-J         $ & $ 0.399$ & $-0.42\pm0.02$ & $-2.51\pm0.27$ & $ 0.17\pm0.03$ & $0.20$ \\
    $H,J-H         $ & $ 1.618$ & $-0.59\pm0.02$ & $-3.29\pm0.31$ & $ 0.13\pm0.04$ & $0.19$ \\
    $[3.6],V-[3.6] $ & $ 0.071$ & $-0.46\pm0.02$ & $-2.55\pm0.27$ & $ 0.18\pm0.03$ & $0.18$ \\
    $[4.5],K-[4.5] $ & $ 0.918$ & $-0.44\pm0.02$ & $-2.59\pm0.29$ & $ 0.19\pm0.04$ & $0.18$ \\
    \multicolumn{6}{c}{\emph{Three-band PW Relations$^b$}} \\
    $B,U-R         $ & $ 1.878$ & $-0.20\pm0.03$ & $-1.95\pm0.33$ & & $0.23$ \\
    $V,B-I         $ & $ 1.365$ & $-0.35\pm0.02$ & $-2.85\pm0.29$ & & $0.19$ \\
    $R,B-I         $ & $ 1.138$ & $-0.39\pm0.02$ & $-2.95\pm0.31$ & & $0.19$ \\
    $I,V-K         $ & $ 0.673$ & $-0.46\pm0.03$ & $-2.99\pm0.33$ & & $0.20$ \\
    $J,V-[3.6]     $ & $ 0.306$ & $-0.44\pm0.03$ & $-2.88\pm0.35$ & & $0.21$ \\
    $K,I-[3.6]     $ & $ 0.220$ & $-0.49\pm0.03$ & $-3.10\pm0.34$ & & $0.20$ \\
    $[3.6],I-[4.5] $ & $ 0.123$ & $-0.47\pm0.03$ & $-2.92\pm0.38$ & & $0.20$ \\
    $H,J-K         $ & $ 1.041$ & $-0.55\pm0.02$ & $-3.39\pm0.31$ & & $0.20$ \\

    \multicolumn{6}{c}{\emph{Three-band PWZ Relations$^c$}} \\
    $B,U-R         $ & $ 1.878$ & $-0.20\pm0.02$ & $-1.55\pm0.30$ & $ 0.14\pm0.04$ & $0.21$ \\
    $V,B-I         $ & $ 1.365$ & $-0.35\pm0.02$ & $-2.59\pm0.27$ & $ 0.11\pm0.04$ & $0.18$ \\
    $R,B-I         $ & $ 1.138$ & $-0.39\pm0.02$ & $-2.68\pm0.29$ & $ 0.11\pm0.04$ & $0.18$ \\
    $I,V-K         $ & $ 0.673$ & $-0.46\pm0.02$ & $-2.65\pm0.26$ & $ 0.15\pm0.03$ & $0.18$ \\
    $J,V-[3.6]     $ & $ 0.306$ & $-0.43\pm0.02$ & $-2.51\pm0.27$ & $ 0.17\pm0.03$ & $0.19$ \\
    $K,I-[3.6]     $ & $ 0.220$ & $-0.49\pm0.02$ & $-2.72\pm0.28$ & $ 0.16\pm0.03$ & $0.18$ \\
    $[3.6],I-[4.5] $ & $ 0.123$ & $-0.46\pm0.02$ & $-2.56\pm0.27$ & $ 0.18\pm0.03$ & $0.18$ \\
    $H,J-K         $ & $ 1.041$ & $-0.55\pm0.02$ & $-3.05\pm0.30$ & $ 0.14\pm0.03$ & $0.18$ \\
\hline
    \multicolumn{5}{p{.9\linewidth}}{NOTE - Only select relations are shown here, the full version of this table is available online}\\
    \multicolumn{5}{p{.9\linewidth}}{$^a$ Color coefficient for the Wesenheit magnitude} \\
    \multicolumn{5}{p{.9\linewidth}}{$^b$ $M = a + b(\log P_F+0.3)$}\\
    \multicolumn{5}{p{.9\linewidth}}{$^c$ $M = a + b(\log P_F +0.3) + c([Fe/H]+1.36)$}
    \end{tabular}
\end{table}
 
\begin{figure}
\includegraphics[scale=0.58]{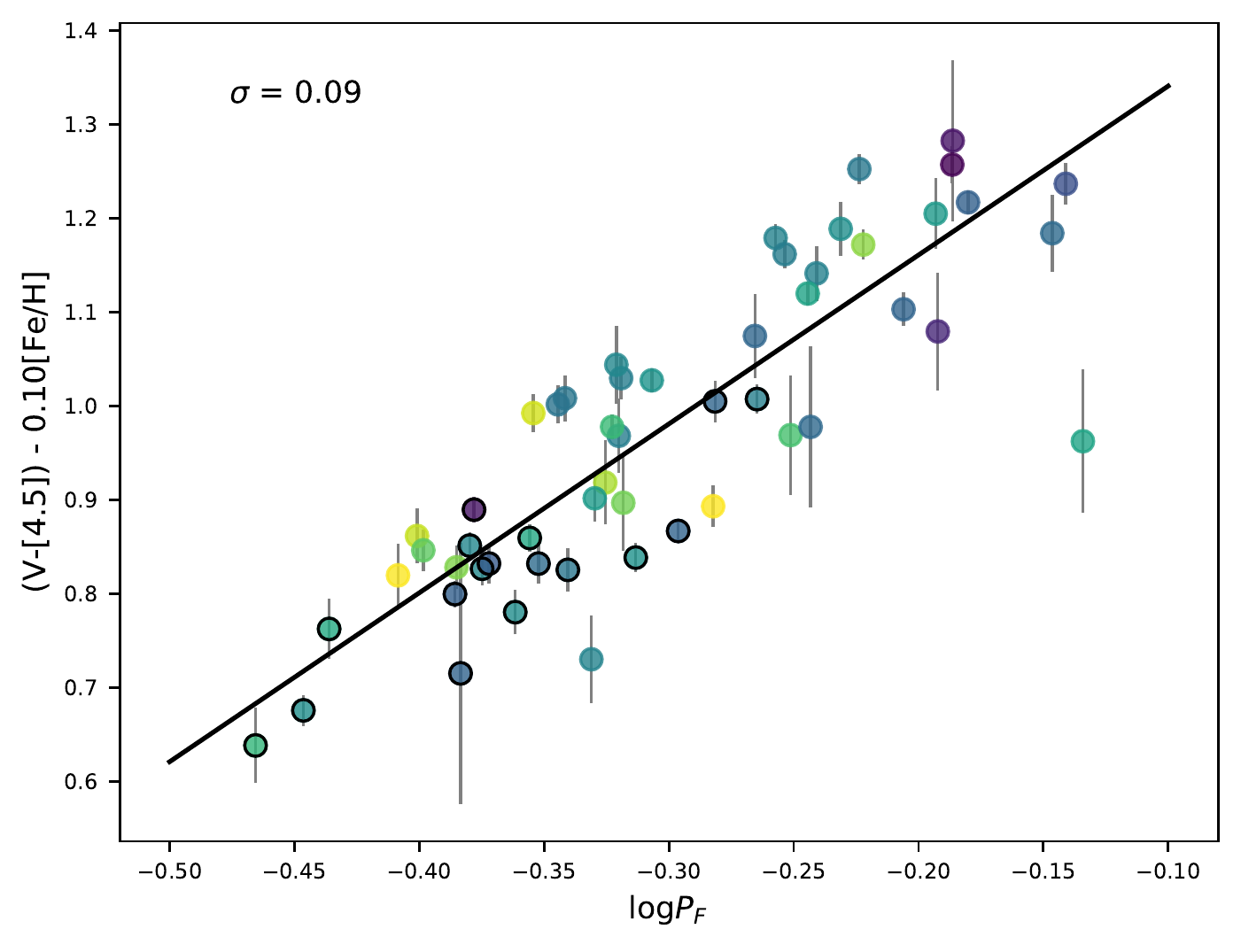}
\centering
\caption{The period-color-metallicity relation using the $(V-[4.5])$ color. The points are color-coded by metallicity. Periods of RRc stars have been fundamentalized and are outlined with black circles. The observed scatter in the PCZ relation is much smaller than the PWZ and PLZ relations, since it is independent of distance.}
\label{fig:pcz} 
\end{figure}

\subsection{Scatter in PLZ relation}\label{sec:scatter}
The scatter around our PWZ and PLZ relations is larger than expected, and only moderately decreases with wavelength. Additionally, the residuals from different passbands are highly correlated. These two observations suggest that there is an unaccounted for source of uncertainty affecting our results that is wavelength independent, and distance and/or metallicity uncertainties are the likely source that could produce the achromatic effect we see here. \protect\citet{lindegren_gaia_2018} states that the formal errors of the parallax in DR2 could be underestimated by up to 30\%, and we are likely seeing this effect here. 

To demonstrate this, we show a period-color-metallicity (PCZ) relation using the $V$ and $[4.5]$ bands in Fig.~\ref{fig:pcz}. This relation is completely independent of distance, but still includes scatter due to measurement uncertainties (i.e. photometry, extinction, and metallicity) and intrinsic scatter (temperature width of the instability strip and evolutionary status of the RRL in our sample). We find $\sigma = 0.09$ mag in the $(V - [4.5])$ PCZ relation, which is more than two times smaller than the equivalent PWZ relation in these bands. This rules out metallicity, extinction, and evolutionary state, because those terms are included in the PCZ. Thus, we conclude that parallax uncertainty, the only term absent from the PCZ, is the primary driver of the increased scatter in the PWZ and PLZ relations.

\begin{figure}
\includegraphics[scale=0.6]{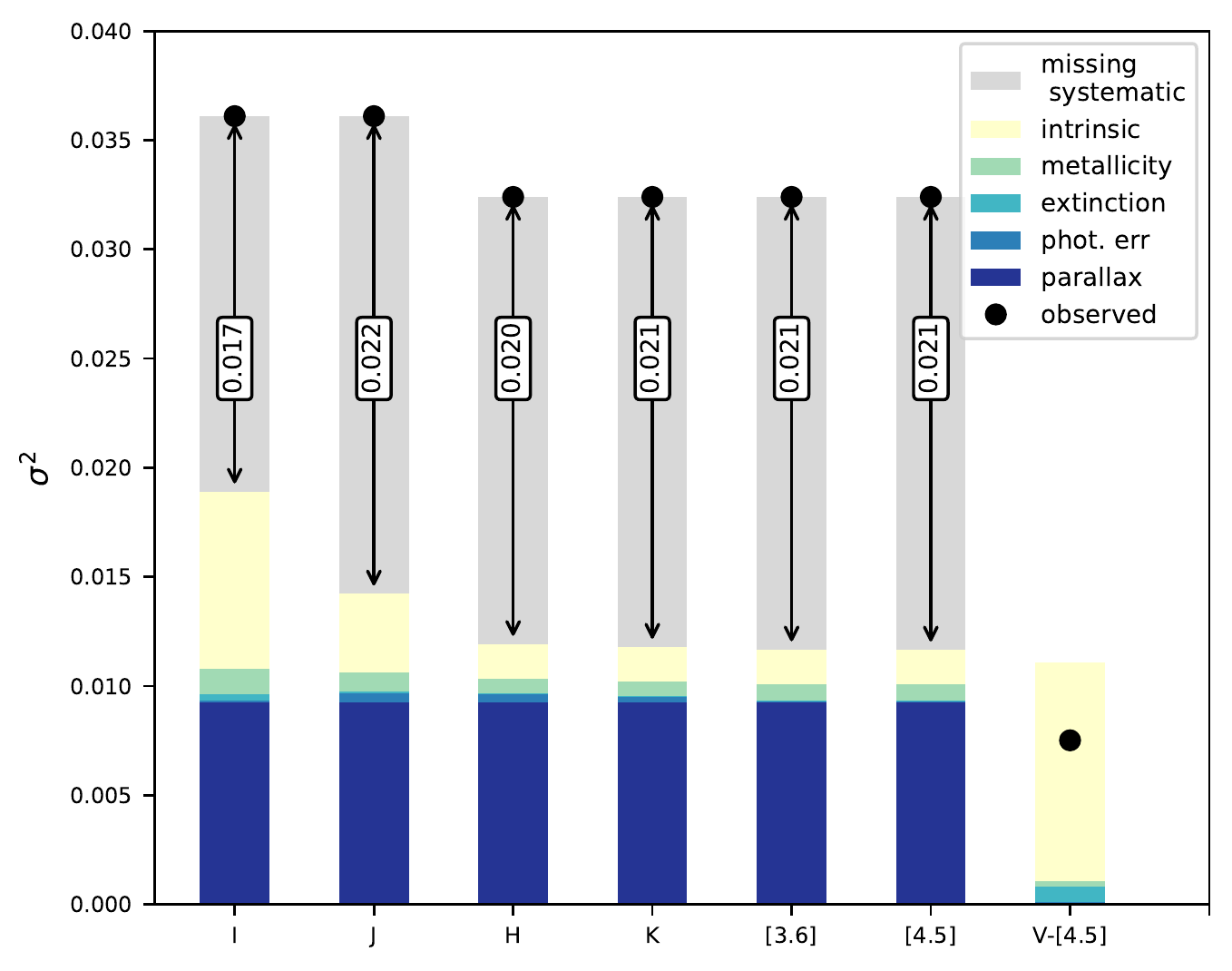}
\centering
\caption{A visual depiction of our error budget for the PLZ relations compared to the PCZ relation. Photometric uncertainties and extinction are barely distinguished, and therefore do not significantly contribute to the total dispersion. For the PLZ relations, the observed variance (black circles) is larger than the quadrature sum of the measurement and astrophysical sources of uncertainty. The total uncertainty (specifically the intrinsic dispersion) in the PCZ relation however is overestimated. This combined with the fact that the missing variance is virtually wavelength independent, we can conclude that we are missing an additional 0.13 mag of dispersion (0.017 mag in variance) in the uncertainties for the distance moduli.}
\label{fig:dispersion} 
\end{figure}

Further evidence of the underestimated parallax uncertainties can be seen in our error budget. In Fig.~\ref{fig:dispersion}, we plot the total observed variance in the PLZ in each band, and in one PCZ relation (black circles). Then, we break down the total variance into identified sources of uncertainty that were described in Section \ref{sec:obs}. Measurement uncertainties include the photometric error (the average uncertainty in the mean magnitude, see Paper I, for each band), distance modulus ($\sigma_{\mu} = 5\times0.434\times \langle \sigma_{\varpi}/\varpi \rangle $), extinction ($\sigma_{A_{\lambda}} = 0.16\langle A_V \rangle \times A_{\lambda}/A_V$), and metallicity ($\sigma_{[Fe/H]} = 0.15 c_{\lambda}$). The intrinsic uncertainty can be attributed to other astrophysical effects, such as the temperature width of the instability strip and deviations in luminosity due to the off-ZAHB evolutionary status of a given RRL in the sample. We adopted the dispersion from the theoretical PLZ relations as the intrinsic dispersion \protect\citep{marconi_new_2015, neeley_new_2017}. After adding in quadrature all our identified sources of uncertainty, we are still well below the observed dispersion in the PLZ for all bands. In the PCZ relation, however, summing the sources of uncertainty actually \emph{overestimates} the observed dispersion, which supports the claim that the parallax uncertainty is underestimated. From the difference between the observed dispersion in the PLZ and identified sources of uncertainty, we calculate that we are missing an average of $\sigma = 0.14$ mag of dispersion. We note that even a 30\% increase in the parallax errors suggested by \protect\citet{lindegren_gaia_2018} does not fully account for the large dispersion we see (there is still 0.12 mag unaccounted for). However, without a more solid handle on the parallax zero point and intrinsic scatter, we cannot provide a reliable estimate of the true errors at this time.

\subsection{Effect of the Gaia DR2 zero-point offset}\label{sec:zp}

\begin{figure}
\includegraphics[scale=0.6]{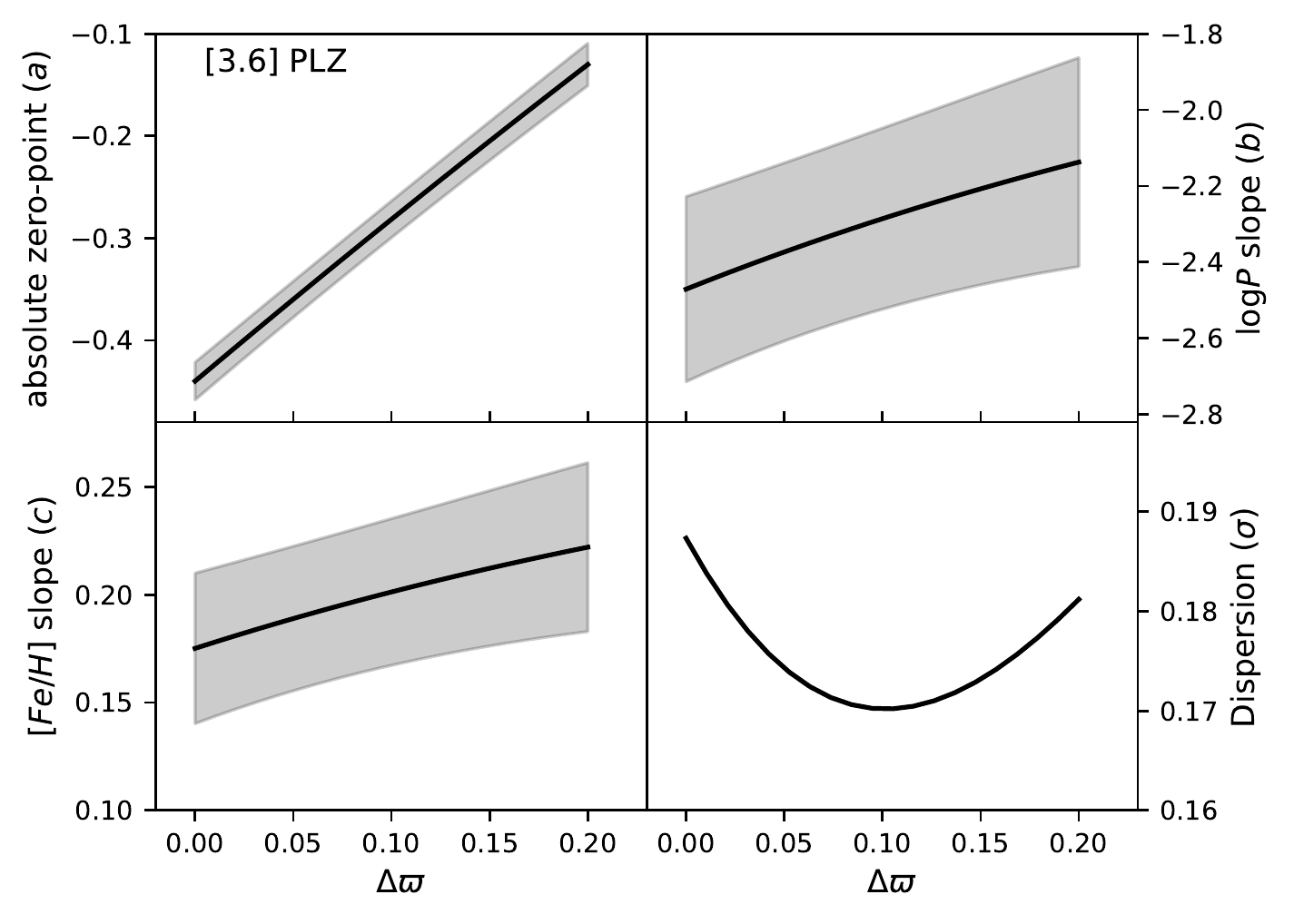}
\centering
\caption{A demonstration of how adjusting the \emph{Gaia} parallax zero-point impacts the terms in the PL(Z). For the $[3.6]$ band, we plot the coefficients of the PLZ relation (top left: absolute zero-point,  top right: period slope, bottom left: metallicity slope, bottom right: dispersion) as a function of Gaia parallax zero-point, $\Delta \varpi$. The shaded regions represent the 1$\sigma$ uncertainties for each parameter. }
\label{fig:plz-off} 
\end{figure}

For our sample of relatively nearby stars, the global parallax zero-point offset discussed in Section~\ref{sec:dr2} can have a systematic impact on the PL(Z). The effect of adopting values of $\Delta \varpi$ between 0 and 0.2 mas on terms for the $[3.6]$ PLZ relation is shown in Fig.~\ref{fig:plz-off}. Larger parallax offsets result in shallower period and steeper metallicity slopes for all wavelengths, but the slopes for different offsets are all consistent within one sigma (the grey band in Fig.~\ref{fig:plz-off}). The PLZ and PWZ zero-points, however, are significantly affected by the parallax offset, which has important implications on distance measurements. For example, when applying the PWZ relation to the LMC, adopting a $\Delta \varpi = 0.06$ mas will return a distance modulus 0.1 mag smaller than if we assumed no offset (a 5\% effect in distance). However, this offset is of the same order of magnitude as the systematic uncertainty of these measurements, and they are therefore still in agreement. This is a smaller effect than what is observed for Classical Cepheids, where the difference in distance moduli was found to be 0.2 mag for the same parallax offsets \protect\citep{groenewegen_cepheid_2018}, but since we are sampling a different range of magnitudes and parallaxes, it is not surprising we find a different result. We would like to emphasize though, that adopting a larger $\Delta \varpi$ will result in measuring systematically smaller distance moduli, and therefore relative distances are unaffected. We also note that an offset of $\Delta \varpi = 0.106$ mas minimizes the dispersion in our sample in most bands, but this value is at the maximum systematic level reported by \emph{Gaia} \protect\citep{lindegren_gaia_2018}. Other authors have reported an offset as large as $0.08$ mas, for a similar range of G magnitudes and distances as the stars in our sample \protect\citep{stassun_evidence_2018}.   

Interestingly, we find the opposite effect to that reported in \protect\citet{muraveva_rr_2018}. They provided PLZ relations in the $K$ and $W1$ (similar to $[3.6]$) bands for two different parallax zero-points, and found that the period slope was steeper and the metallicity slope shallower for the case of the larger zero-point offset. Both their study and ours report a smaller zero-point for the larger offset. While the direction of the effect may be different, the individual slopes are still consistent between the two methods for adopting the same offset, and so it is possible the difference can be attributed to statistical fluctuations.

\section{Distance Measurements}\label{sec:dist}

In this section, we test the performance of our PL(Z) and PW(Z) relations as distance indicators. We provide an estimate of the precision anticipated given certain conditions, and measure the distance to three well-studied systems, the Galactic globular cluster M4, and the Large and Small Magellanic Clouds. 

\subsection{Summary of distance uncertainties}

As we wait for better parallaxes in \emph{Gaia} DR3, it is useful to assess the current precision to which we can measure distances to individual stars. This helps give an idea of the current limits of our relations in practice, given realistic observing constraints such as a lack of abundance measurements, or photometry in a limited number of passbands.  

\begin{figure}
\includegraphics[scale=0.8]{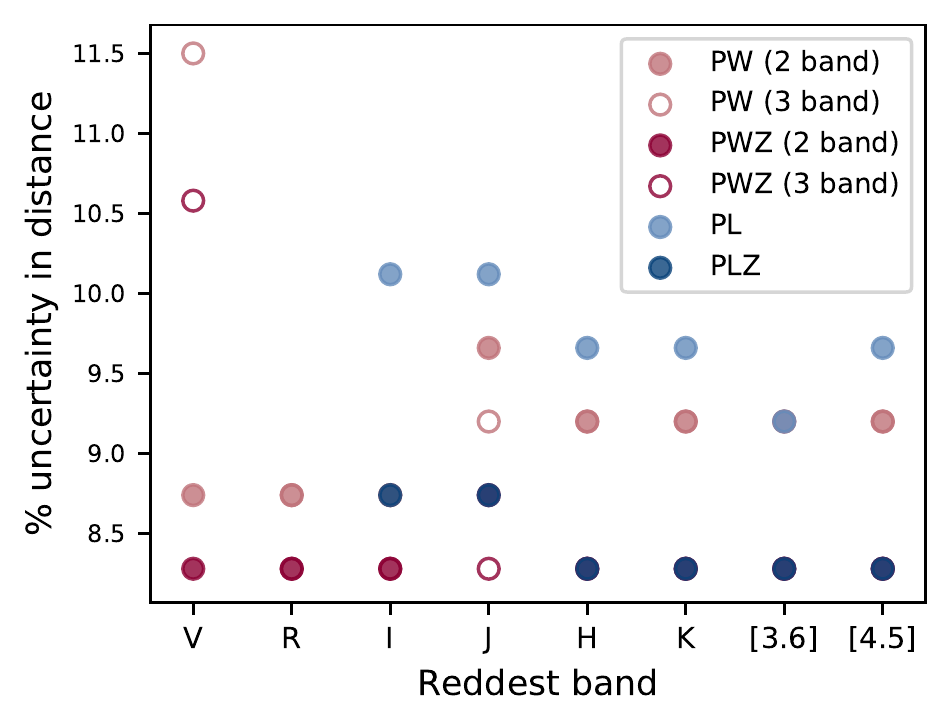}
\centering
\caption{An estimate of the expected percent uncertainty in distance determined for an individual star for various relations determined in this paper. Each point represents the smallest possible dispersion using that type of relation, where the reddest passband used is indicated on the x-axis. In general, PLZ relations offer just over $1\%$ more precision than PLs, and PWZ relations are comparable to PW in the purely optical combinations, but $\sim 1\%$ more precise when using infrared bands. PWZ and PLZ relations offer similar precision. }
\label{fig:dist-err} 
\end{figure}

To estimate the precision of a given relation, we simply project the observed dispersion to a distance uncertainty ($\sigma_d = 0.46 \sigma_{\mu}$). This estimate does not include other statistical or systematic effects in the measurement (i.e. extinction, measurement uncertainties, or parallax zero-point). In Fig.~\ref{fig:dist-err}, we show the distance uncertainty expected for a given band, where that band is defined as the reddest band in the observations. For example, for the $H$ band, we plot the PW relations for $W(H,B-H)$ and $W(R,B-H)$, and the PWZ relations for $W(H,B-H)$ and $W(I,B-H)$, because these combinations have the smallest dispersion. We have excluded any combination using the \emph{Gaia} bands, to compare all relations using the same set of calibrators. 

Using the results of this work, the inclusion of metallicity generally improves distance precision by $1\%$. For the $J$ and longer bands, there is no improvement in precision using a PWZ versus a PLZ, and the advantage of the PWZ is only the reduced sensitivity to extinction. Three-band PWZ relations also tend to have a slight advantage over two band combinations. We expect the uncertainty of RRL distance measurements to be drastically reduced as more accurate parallaxes become available. We also expect the improvement in precision between PW and PWZ (or PL and PLZ) relations to become more apparent.

\subsection{Globular Cluster M4}

\begin{figure}
\includegraphics[scale=0.9]{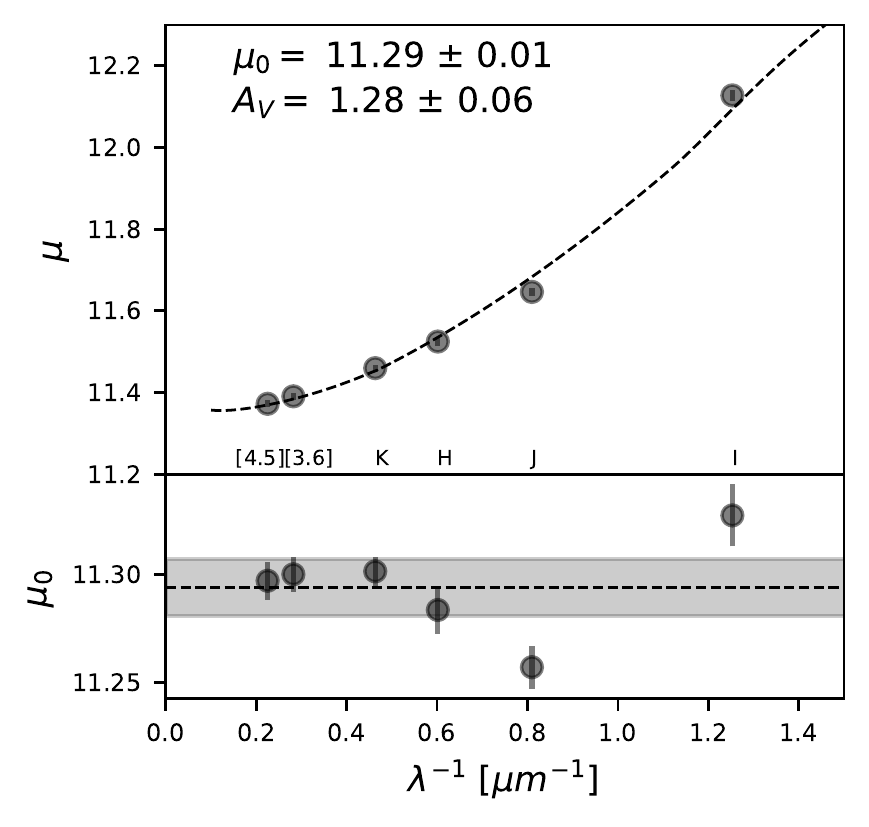}
\centering
\caption{\emph{Top}: Reddened distance moduli versus $\lambda^{-1}$ for M4 from $I$ to $[4.5]$. The dashed line is the Cardelli reddening law fit to the data with an $R_V = 3.62$, as suggested in \protect\citet{hendricks_new_2012}, and the zero-point of this line is the true distance modulus. \emph{Bottom}: The distance modulus derived for each passband has now been corrected for the best fit extinction. The uncertainty for each passband is the standard error of the mean distance modulus derived from individual stars in the cluster. The shaded region represents the $1\sigma$ uncertainty in the mean. }
\label{fig:M4} 
\end{figure}

For M4, we adopt the same approach used in \protect\citet{neeley_new_2017} to fit the distance modulus and extinction simultaneously. Multi-wavelength average magnitudes of the RRL in M4 are available in the literature \protect\citep{stetson_optical_2014, neeley_distance_2015}, and we use our PLZ relations (from $I$ to $[4.5]$) to compute a reddened distance modulus for each passband. The true distance modulus and visual band extinction, $A_V$, were then derived by fitting the reddened distance moduli to the Cardelli reddening law with an $R_V = 3.62$, as suggested by  \protect\citet{hendricks_new_2012}. We obtain $\mu_0 = 11.29 \pm 0.01 (stat) \pm 0.02 (syst)$  mag and $A_V = 1.28 \pm 0.06$ mag. The statistical uncertainty is the uncertainty in the fit, whereas the systematic uncertainty is the uncertainty in the mid-infrared zero-point. Fig.~\ref{fig:M4} shows the distance modulus derived in each band, and corrected using the best-fit extinction. These results are in excellent agreement with the results from \protect\citet{hendricks_new_2012} ($\mu_0 = 11.28\pm0.06$ mag and $A_V = 1.39\pm0.07$ mag), as well as the distance modulus measured via eclipsing binaries, $\mu_0 = 11.30\pm 0.05$ mag \protect\citep{kaluzny_cluster_2013}. \emph{HST} calibrations of the PL relation tend to result in systematically higher values, \protect\citep[i.e. $\mu_{0,[3.6]} = 11.353\pm0.095$ mag from][]{neeley_new_2017}, but are still within $1\sigma$ of our current result.

\begin{figure*}
\includegraphics[scale=0.9]{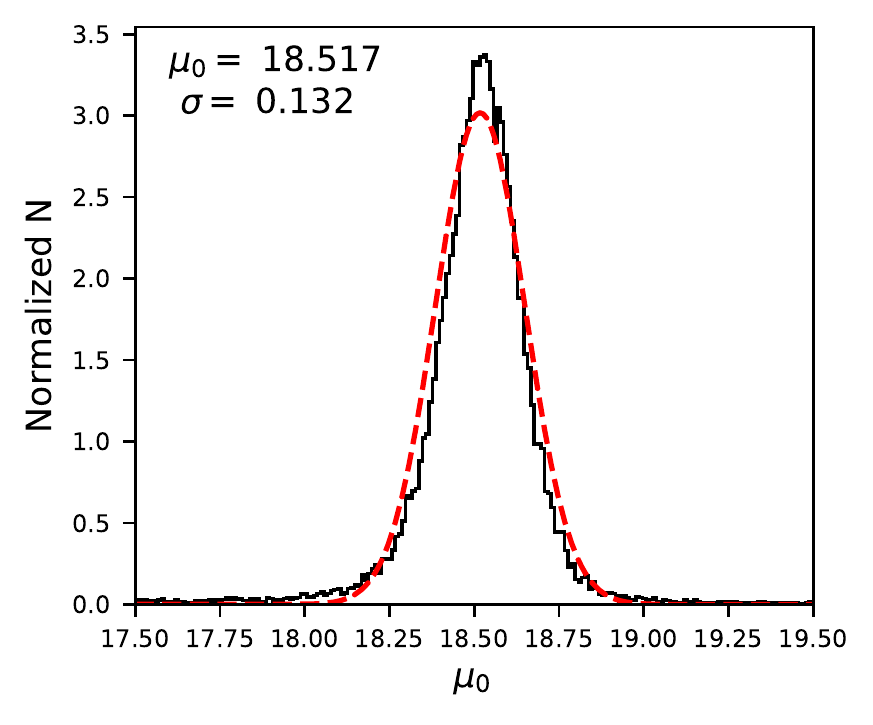}
\includegraphics[scale=0.9]{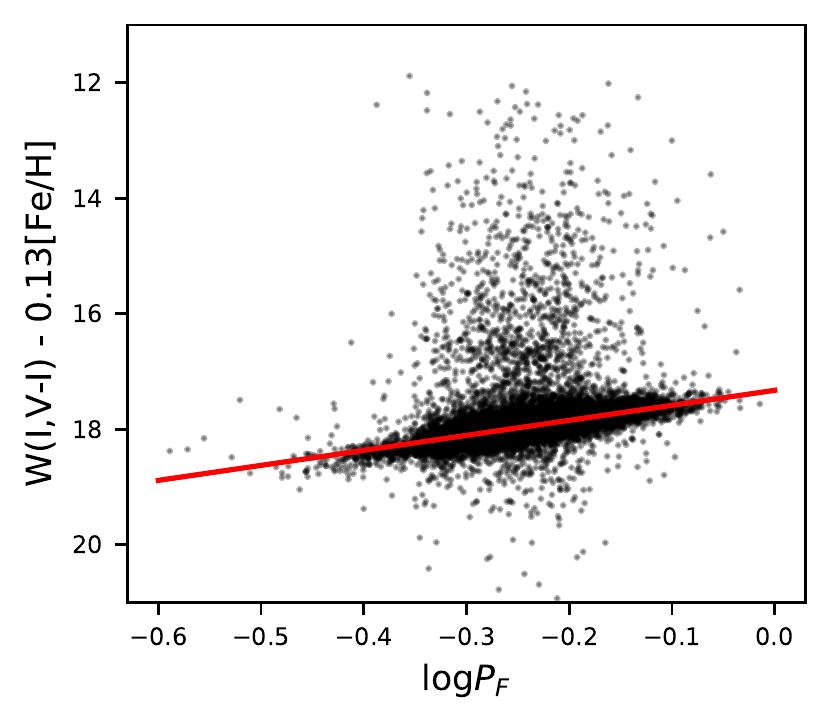}
\includegraphics[scale=0.9]{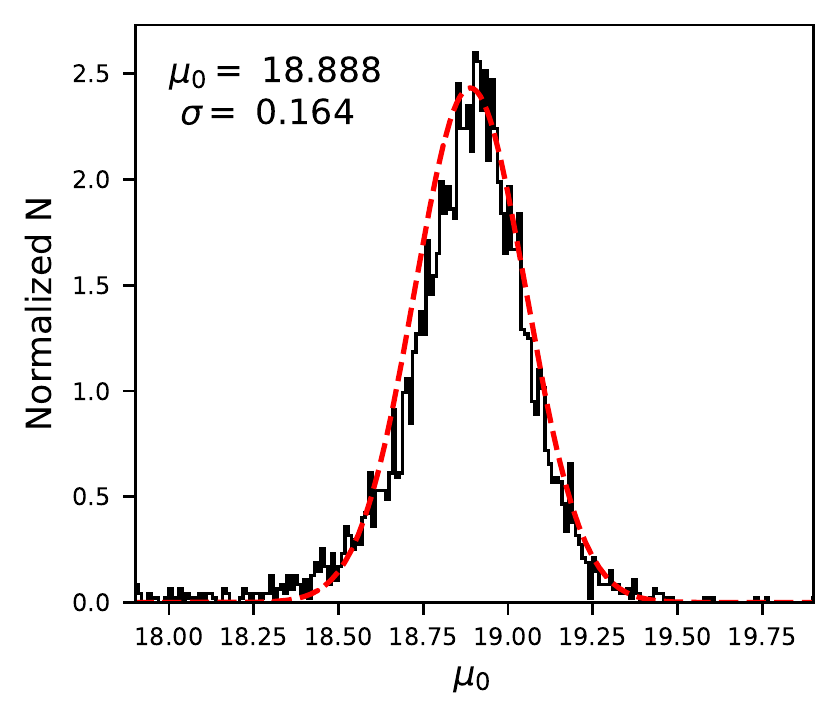}
\includegraphics[scale=0.9]{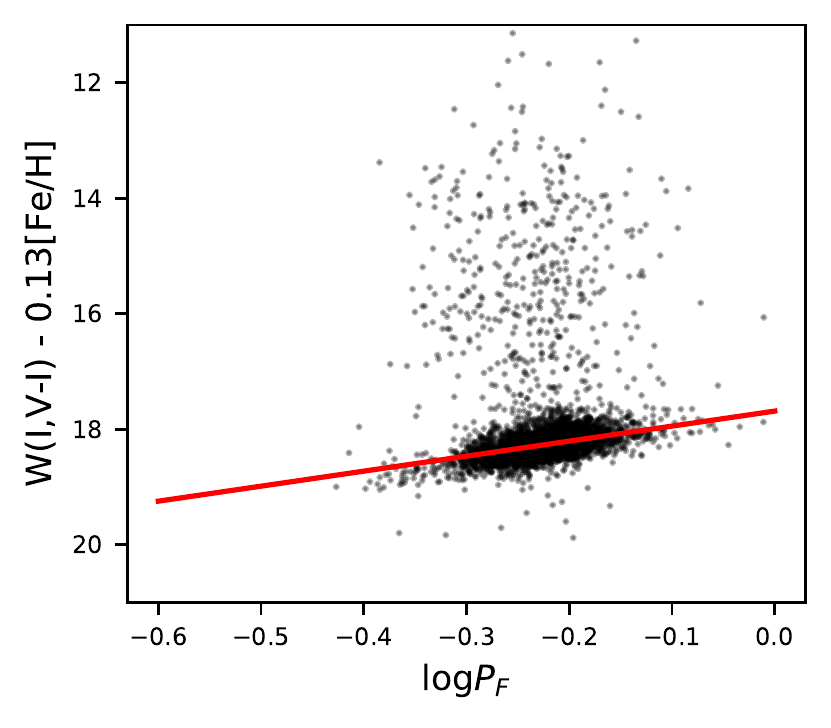}
\centering
\caption{\emph{Top left}: A histogram of the true distance moduli for individual stars in the LMC, computed using a period-Wesenheit-metallicity relation. The red dashed line is a Gaussian fit, which was used to derive the mean distance modulus to the LMC. The quoted sigma, measured from the width of the Gaussian, is a combination of the line-of-sight depth of the LMC and the statistical uncertainties in the distance estimate. \emph{Top right}: The Wesenheit relation for all stars in the OGLE LMC catalog. Milky Way foreground stars have not been removed, and are clearly identified as those stars between 12 and 17 mag. The red line is the Wesenheit relation measured in this work using Galactic field stars, shifted by the mean distance modulus derived from the histogram. \emph{Bottom row}: Same as the top left panel but for the SMC.   }
\label{fig:xmc} 
\end{figure*}

\subsection{Magellanic Clouds}
The Magellanic Clouds provide another important test for our PLZ relations, because they are commonly adopted as anchors for the extragalactic distance scale. For the RRL in the Magellanic Clouds, fewer passbands are available, and we must take a different approach to measure their distances than in M4. Mean $V$ and $I$ band magnitudes of 35,492 and 5,671 RRL in the Large Magellanic Cloud (LMC) and Small Magellanic Cloud (SMC), respectively, are available through the Optical Gravitational Lensing Experiment (OGLE) Collection of Variable Stars \protect\citep{soszynski_ogle_2016}. The metallicity of each RRab star was estimated from its Fourier components (provided in the OGLE catalog), using Equation 2 from \protect\citet{smolec_metallicity_2005}, transformed into the \protect\citet{zinn_globular_1984} metallicity scale. Since there is currently no $I$ band calibration to derive metallicity from the Fourier parameters for RRc stars (although a $V$ band calibration is available), we do not include the RRc stars in further analysis. We find an average metallicity of $\langle [Fe/H] \rangle = -1.5$ and $\langle [Fe/H] \rangle = -1.7$ dex for the LMC and SMC respectively.  

The distances to individual stars in the LMC and SMC are determined from our PWZ relation, using the Wesenheit magnitude $W(I,V-I)$ and the individual metallicities calculated above (Fig.~\ref{fig:xmc}). We do note that the dust properties vary significantly across the LMC and SMC \protect\citep{gordon_quantitative_2003}, but adopting a constant $R_V = 3.1$ is a reasonable approximation for this work. The mean distance was then determined from the peak of the resulting distribution of distances to avoid introducing biases from the bright foreground stars (Fig.~\ref{fig:xmc}). For the LMC, we measure $\mu_0 = 18.517\pm 0.001(stat)\pm0.104(syst)$ mag, and for the SMC we measure $\mu_0 = 18.888\pm0.002(stat)\pm 0.104 (syst)$ mag. A summary of our error budget is available in Table~\ref{tab:errs}. For the statistical uncertainty, we assumed conservative estimates of 0.02 mag for the uncertainty of the mean OGLE $V$ and $I$ magnitudes, which translates to 0.05 mag when propagating to the Wesenheit magnitude. Similarly, we assumed 0.02 mag for the uncertainty of the photometric calibration of the OGLE $V$ and $I$ bands. The remaining terms come directly from the PWZ relation, and we assumed 20\% uncertainties on our individual metallicity values. The statistical uncertainty has been divided by the square root of the number of stars used in the fit, 23957 in the LMC and 4233 in the SMC. We do note however, that we are not currently accounting for the systematics in the \emph{Gaia} parallax zero-point, but this is beyond the scope of the paper considering these distance measurements are only meant to be a consistency check of our relations. 

\begin{table}
    \scriptsize
    \centering
    \caption{Summary of the error budget for distance measurement of the Magellanic Clouds}
    \label{tab:errs}
    \begin{tabular}{l c c c}
    \hline
    Description & Term & LMC & SMC  \\
    \hline
    \multicolumn{4}{c}{\emph{Statistical Uncertainties (per star)}} \\
    Mean Wesenheit magnitude & $\sigma_{m_W}$ & 0.05 & 0.05 \\
    PWZ period slope & $\sigma_{b}|\log P_F|$ & 0.06 & 0.06 \\
    PWZ metallicity slope & $c[Fe/H]\sqrt{\big(\frac{\sigma_c}{c}\big)^2 + \big(\frac{\sigma_{[Fe/H]}}{[Fe/H]}\big)^2}$ & 0.06 & 0.07 \\
    \multicolumn{4}{c}{\emph{Systematic Uncertainties}} \\ 
    OGLE phot. calibration & $\sigma_{ZP_W}$ & 0.05 & 0.05 \\
    Zero-point & $\sigma_a$ & 0.02 & 0.02 \\ 
    PWZ period constant & $0.3\sigma_b$ & 0.08 & 0.08 \\ 
    PWZ metallicity constant & $1.36\sigma_{[Fe/H]}$ & 0.04 & 0.04 \\ 
    \hline
    Total Statistical Uncertainty & & 0.001 & 0.002 \\
    Total Systematic Uncertainty & & 0.104 & 0.104 \\
    \hline
    \end{tabular}
\end{table}

The values of the distance moduli we measure are in excellent agreement with the many estimates from the literature, including estimates based on the Cepheid Leavitt law,  \protect\citep[$\mu_{0,LMC} = 18.48 \pm 0.03$ and $\mu_{0,SMC} = 18.96\pm0.01\pm0.03$,][respectively]{freedman_carnegie_2012, scowcroft_carnegie_2016}, and eclipsing binaries \protect\citep[$\mu_{0,LMC} = 18.477\pm0.004(stat)\pm0.026(syst)$;][]{pietrzynski_distance_2019}. Our estimates also agree well with the compendium and consensus values obtained for each galaxy, $\mu_{0,LMC}=18.49\pm0.09$ mag \protect\citep{de_grijs_clustering_2014} and $\mu_{0,SMC}=18.96\pm0.02$ mag \protect\citep{de_grijs_clustering_2015}. 

\section{Conclusions}

We have presented new calibrations of the RRL PW(Z) and PL(Z) relations using the CRRP sample. The effect of metallicity can clearly be seen, where more metal-rich stars tend to have fainter absolute magnitudes than metal-poor stars of the same period. Additionally, metallicity seems to have a larger effect at optical wavelengths, which is not predicted by current theoretical models. 

The remaining scatter ($\sim0.18$ mag in the infrared) in all of our relations is larger than expected from previous theoretical and empirical results, but can be plausibly attributed to presently unaccounted uncertainties and/or systematics in the distances. This is supported by the strong (achromatic) correlation of the residuals for individual stars observed at different wavelengths, and corroborated by the much smaller scatter seen in the distance-independent PCZ relations. 

Using our PLZ and PWZ relations, we determine distances to the globular cluster M4 and the Magellanic Clouds. Our distance for M4, $\mu_0 = 11.29\pm0.01(stat)\pm0.02(syst)$ mag, was obtained by simultaneously fitting the distance and extinction to M4 using multi-wavelength data, and is in excellent agreement with literature estimates. For the Magellanic Clouds, we used a Wesenheit magnitude constructed from the $V$ and $I$ bands to derive distances of RRL in the OGLE database. We also find good agreement with distances derived with the Cepheid Leavitt Law in the infrared and from Eclipsing Binaries.  

The results presented in this paper show that RRL PLZ and PWZ relations can realize the promise of an independent distance indicator with accuracy comparable to the Cepheid Leavitt Law. We anticipate that upcoming \emph{Gaia} data releases with better-understood biases and global zero-point offsets, will allow us to further reduce the current systematic error in the RRL PLZ relations, providing absolute distances with uncertainty dominated by statistical error of the order of a few percent.

\section*{Acknowledgements}
We thank the anonymous referee, whose insightful comments improved the clarity and strength of this work. This work was partially supported by the National Science Foundation under Grant No. AST-1714534. 
This work has made use of data from the European Space Agency (ESA) mission
\href{https://www.cosmos.esa.int/gaia}{Gaia}), processed by the {\it Gaia}
Data Processing and Analysis Consortium (\href{https://www.cosmos.esa.int/web/gaia/dpac/consortium}{DPAC}). Funding for the DPAC
has been provided by national institutions, in particular the institutions
participating in the {\it Gaia} Multilateral Agreement.
This work is based [in part] on observations made with the Spitzer Space Telescope, which is operated by the Jet Propulsion Laboratory, California Institute of Technology under a contract with NASA.
Support for this work was provided by NASA through Hubble Fellowship grant \#51386.01 awarded to to R.L.B. by the Space Telescope Science Institute, which is operated by the Association of Universities for Research in Astronomy, Inc., for NASA, under contract NAS 5-26555.


\bibliographystyle{mnras}



\appendix

\section{Alternate Fitting Techniques}

Traditional linear regressions assume that there is no uncertainty in the independent variables, but this is not the case for our data set. While the uncertainties on the period can be considered negligible ($\sigma_P \approx 10^{-5}$ days), the metallicity uncertainties cannot. Additionally, there is a reasonable concern that our relations could be biased, particularly given our small sample size, by a few stars with absolute magnitudes consistently brighter than stars of a similar period. Therefore, we explored two parallel analyses, which are described here.

\subsection{Robust Analysis}
To explore the influence of outlying data points, we followed the same procedure detailed in the main text, but replaced the weighted least squares fit with a robust linear regression (RLM in the Python statsmodels package). The robust method iteratively re-weights data points based on their median absolute deviations from the fitted line. The coefficients derived using the robust analysis are available in Table~\ref{tab:robust}. We find period slopes that are consistently steeper, but generally within $1\sigma$ of the weighted least squares analysis, and metallicity slopes that are slightly steeper but also consistent with the results in the main text. Using these relations, we derive distances of $\mu_0 = 11.31\pm0.01$, $\mu_0 = 18.558\pm0.001(stat)\pm0.104(syst)$, and $\mu_0 = 18.941\pm0.002(stat)\pm0.104(syst)$ for M4, the LMC and SMC respectively.

\begin{table}
    \scriptsize
    \centering
    \caption{Robust Fitting Results}
    \label{tab:robust}
    \begin{tabular}{l c cccc}
    \hline 
    Filters & $\alpha^a$ & a & b & c & $\sigma$ \\
    \hline
    \multicolumn{6}{c}{\emph{PL Relations$^b$}} \\
    $I$     & &  $0.14\pm0.03$ & $-2.39\pm0.38$ & & 0.22 \\
    $J$     & & $-0.16\pm0.03$ & $-2.76\pm0.35$ & & 0.22 \\
    $H$     & & $-0.33\pm0.03$ & $-3.08\pm0.32$ & & 0.21 \\
    $K_s$   & & $-0.39\pm0.03$ & $-3.16\pm0.33$ & & 0.20 \\
    $[3.6]$ & & $-0.41\pm0.02$ & $-3.21\pm0.35$ & & 0.20 \\
    $[4.5]$ & & $-0.43\pm0.03$ & $-3.25\pm0.36$ & & 0.20 \\
    \multicolumn{6}{c}{\emph{PLZ Relations$^c$}} \\
    $I$     & & $0.14\pm0.03$ & $-1.90\pm0.29$ & $0.23\pm0.04$ & 0.18 \\
    $J$     & & $-0.17\pm0.03$ & $-2.24\pm0.27$ & $0.21\pm0.04$ & 0.19 \\
    $H$     & & $-0.33\pm0.02$ & $-2.65\pm0.24$ & $0.19\pm0.04$ & 0.18 \\
    $K_s$   & & $-0.39\pm0.02$ & $-2.73\pm0.25$ & $0.18\pm0.04$ & 0.18 \\
    $[3.6]$ & & $-0.41\pm0.02$ & $-2.79\pm0.24$ & $0.19\pm0.04$ & 0.17 \\
    $[4.5]$ & & $-0.43\pm0.02$ & $-2.82\pm0.25$ & $0.19\pm0.04$ & 0.18 \\
    \multicolumn{6}{c}{\emph{Two-band PW Relations$^b$}} \\
    $B,U-B$          & $ 6.228$ & $ 0.23\pm0.06$ & $ 0.98\pm0.67$ &  & $0.40$ \\
    $V,B-V$          & $ 3.058$ & $-0.28\pm0.03$ & $-3.22\pm0.27$ &  & $0.19$ \\
    $R,B-R$          & $ 1.689$ & $-0.42\pm0.03$ & $-3.47\pm0.29$ &  & $0.19$ \\
    $I,V-I$          & $ 1.467$ & $-0.43\pm0.03$ & $-3.22\pm0.29$ &  & $0.19$ \\
    $J,V-J$          & $ 0.399$ & $-0.44\pm0.03$ & $-3.11\pm0.33$ &  & $0.21$ \\
    $H,J-H$          & $ 1.618$ & $-0.60\pm0.03$ & $-3.66\pm0.32$ &  & $0.20$ \\
    $[3.6],V-[3.6]$  & $ 0.071$ & $-0.48\pm0.03$ & $-3.31\pm0.35$ &  & $0.20$ \\
    $[4.5],K-[4.5]$  & $ 0.918$ & $-0.46\pm0.03$ & $-3.32\pm0.38$ &  & $0.21$ \\
    \multicolumn{6}{c}{\emph{Two-band PWZ Relations$^c$}} \\
    $B,U-B         $ & $ 6.228$ & $ 0.22\pm0.05$ & $ 1.62\pm0.62$ & $ 0.33\pm0.10$ & $0.36$ \\
    $V,B-V         $ & $ 3.058$ & $-0.29\pm0.03$ & $-3.06\pm0.28$ & $ 0.08\pm0.05$ & $0.18$ \\
    $R,B-R         $ & $ 1.689$ & $-0.42\pm0.03$ & $-3.28\pm0.27$ & $ 0.10\pm0.05$ & $0.18$ \\
    $I,V-I         $ & $ 1.467$ & $-0.43\pm0.02$ & $-2.89\pm0.22$ & $ 0.15\pm0.04$ & $0.18$ \\
    $J,V-J         $ & $ 0.399$ & $-0.45\pm0.03$ & $-2.69\pm0.26$ & $ 0.18\pm0.04$ & $0.19$ \\
    $H,J-H         $ & $ 1.618$ & $-0.61\pm0.03$ & $-3.37\pm0.33$ & $ 0.16\pm0.04$ & $0.19$ \\
    $[3.6],V-[3.6] $ & $ 0.071$ & $-0.48\pm0.02$ & $-2.91\pm0.24$ & $ 0.18\pm0.04$ & $0.17$ \\
    $[4.5],K-[4.5] $ & $ 0.918$ & $-0.46\pm0.02$ & $-2.91\pm0.25$ & $ 0.20\pm0.04$ & $0.18$ \\
    \multicolumn{6}{c}{\emph{Three-band PW Relations$^b$}} \\
    $B,U-R         $ & $ 1.878$ & $-0.22\pm0.03$ & $-2.13\pm0.33$ &  & $0.23$ \\
    $V,B-I         $ & $ 1.365$ & $-0.37\pm0.02$ & $-3.22\pm0.28$ &  & $0.19$ \\
    $R,B-I         $ & $ 1.138$ & $-0.40\pm0.02$ & $-3.30\pm0.30$ &  & $0.19$ \\
    $I,V-K         $ & $ 0.673$ & $-0.48\pm0.03$ & $-3.28\pm0.32$ &  & $0.20$ \\
    $J,V-[3.6]     $ & $ 0.306$ & $-0.45\pm0.03$ & $-3.17\pm0.35$ &  & $0.21$ \\
    $K,I-[3.6]     $ & $ 0.220$ & $-0.51\pm0.03$ & $-3.34\pm0.33$ &  & $0.20$ \\
    $[3.6],I-[4.5] $ & $ 0.123$ & $-0.48\pm0.03$ & $-3.32\pm0.35$ &  & $0.20$ \\
    $H,J-K         $ & $ 1.041$ & $-0.57\pm0.03$ & $-3.52\pm0.30$ &  & $0.20$ \\
    \multicolumn{6}{c}{\emph{Three-band PWZ Relations$^c$}} \\
    $B,U-R         $ & $ 1.878$ & $-0.23\pm0.03$ & $-1.81\pm0.31$ & $ 0.17\pm0.05$ & $0.21$ \\
    $V,B-I         $ & $ 1.365$ & $-0.37\pm0.02$ & $-2.96\pm0.25$ & $ 0.12\pm0.04$ & $0.18$ \\
    $R,B-I         $ & $ 1.138$ & $-0.41\pm0.02$ & $-3.06\pm0.25$ & $ 0.12\pm0.04$ & $0.18$ \\
    $I,V-K         $ & $ 0.673$ & $-0.48\pm0.02$ & $-2.91\pm0.24$ & $ 0.16\pm0.04$ & $0.18$ \\
    $J,V-[3.6]     $ & $ 0.306$ & $-0.46\pm0.03$ & $-2.74\pm0.26$ & $ 0.18\pm0.04$ & $0.19$ \\
    $K,I-[3.6]     $ & $ 0.220$ & $-0.51\pm0.02$ & $-2.93\pm0.26$ & $ 0.18\pm0.04$ & $0.18$ \\
    $[3.6],I-[4.5]$ & $0.123$ & $-0.48\pm0.02$ & $-2.91\pm0.24$ & $0.18\pm0.04$ & $0.18$ \\
    $H,J-K$ & $1.041$ & $-0.57\pm0.02$ & $-3.17\pm0.27$ & $0.17\pm0.04$ & $0.18$ \\
\hline
    \multicolumn{5}{p{.9\linewidth}}{NOTE - Only select relations are shown here, the full version of this table is available online}\\
    \multicolumn{5}{p{.9\linewidth}}{$^a$ Color coefficient for the Wesenheit magnitude} \\
    \multicolumn{5}{p{.9\linewidth}}{$^b$ $M = a + b(\log P_F +0.3)$}\\
    \multicolumn{5}{p{.9\linewidth}}{$^c$ $M = a + b(\log P_F +0.3) + c([Fe/H]+1.36)$}
    \end{tabular}
\end{table}

\subsection{Bayesian Analysis}
To test the effect of metallicity uncertainties on our derived relations, we performed a Bayesian analysis that allows for multidimensional uncertainties. The details of this method are outlined in Section 7 of \protect\citet{hogg_data_2010}, and have been generalized to three dimensions following D. Foreman-Mackey\footnote{see \href{http://doi.org/10.5281/zenodo.3221478}{DOI 10.5281}}. We only made two customizations. First, we used a diagonal correlation matrix, since the metalliciy and photometry are independent datasets, and therefore their uncertainties are not correlated. Second, we used a standard Cauchy prior distribution for the slopes $b$ and $c$, which provides a uniform distribution in the slope angles $\theta_1$ and $\theta_2$ (where $b=\tan \theta_1$ and $c=\tan \theta_2$).

\begin{figure}
\includegraphics[scale=0.7]{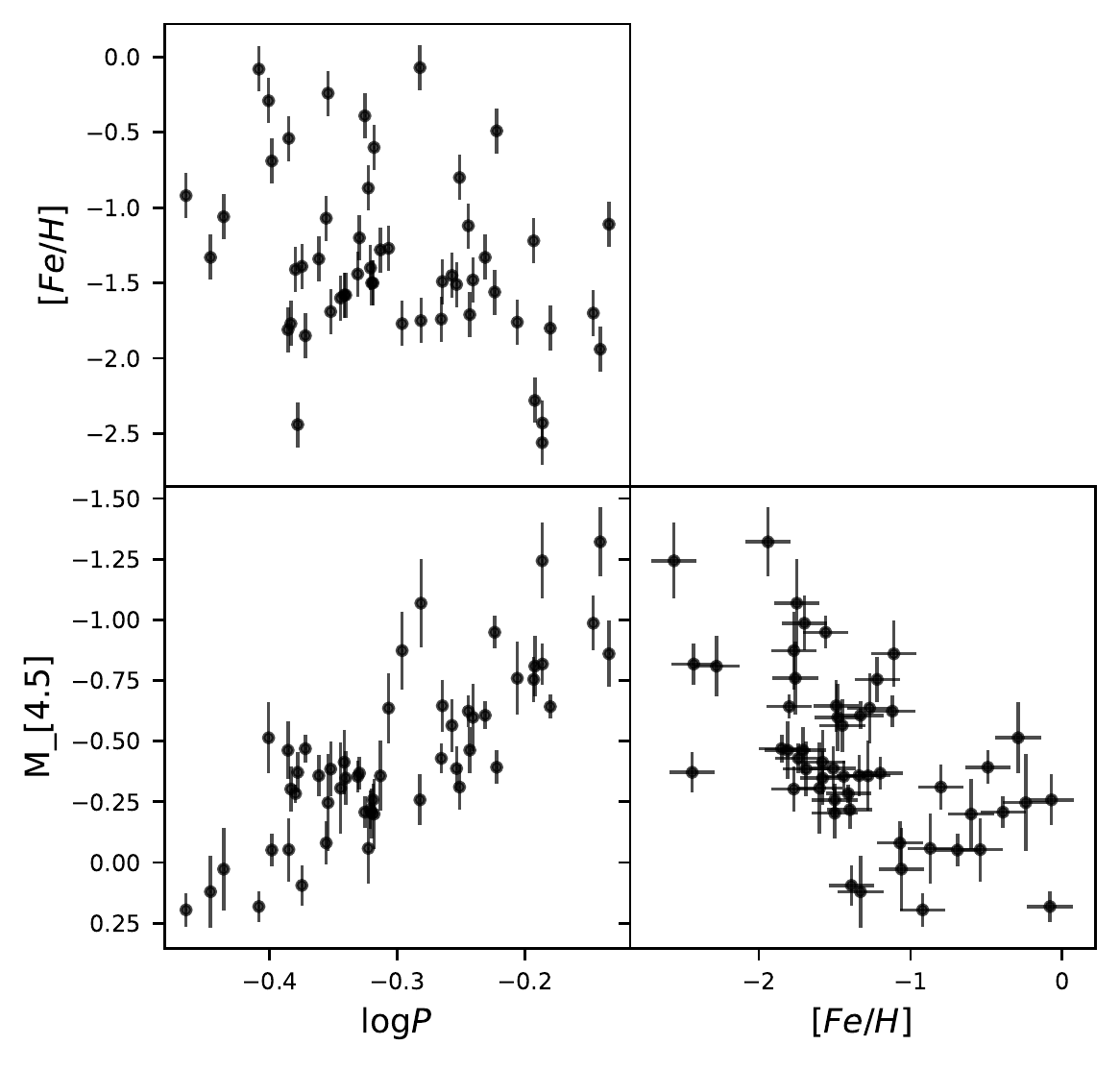}
\centering
\caption{3D distribution of the data in the $[4.5]$ band. The uncertainty in $\log P$ is too small to be visible.}
\label{fig:3d-data} 
\end{figure}

\begin{figure*}
\includegraphics[scale=0.9]{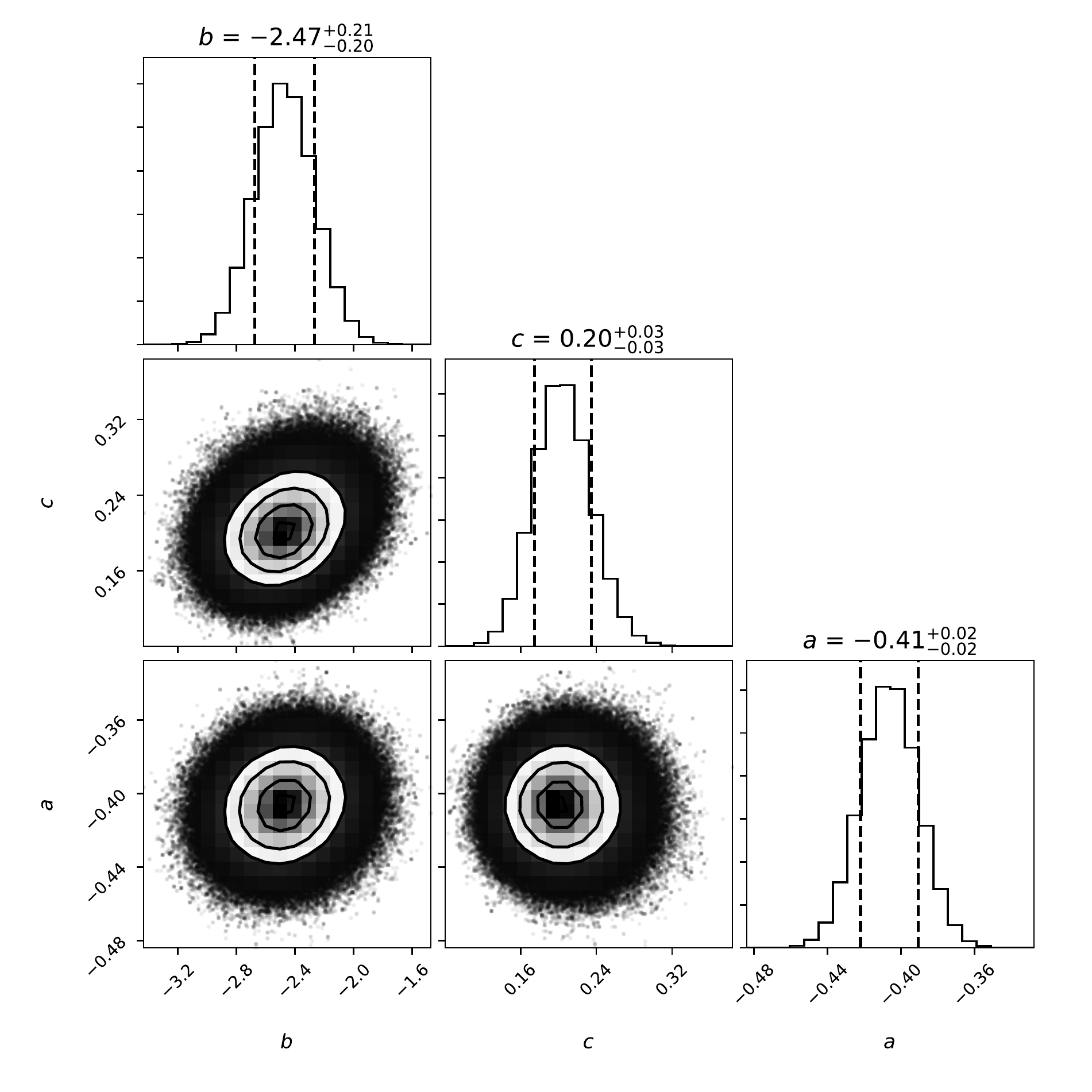}
\centering
\caption{Posterior distribution of the coefficients of the $[4.5]$ band PLZ relation. Their $1\sigma$ uncertainties are indicated by the dashed lines in the histograms.}
\label{fig:3d-results} 
\end{figure*}

In Figure~\ref{fig:3d-data}, we plot the 3D distribution of our data in the $[4.5]$ band. Uncertainty in metallicity is assumed to be 0.15 dex for all stars, since individual uncertainties are not available in \protect\citet{fernley_absolute_1998}. Figure~\ref{fig:3d-results} shows the posterior distribution of the coefficients of the $[4.5]$ band PLZ relation. The derived coefficients are listed in Table~\ref{tab:bayesian}. The period slopes are very consistent with the results of the weighted least squares results presented in Table~\ref{tab:results}, and the metallicity slope tends to be steeper at about the $1\sigma$ level for PLZ relations, and up to $2\sigma$ steeper for the PWZ relations. The steeper metallicity slopes only have a moderate effect on the distances derived in Section~\ref{sec:dist}; using these relations results in $\mu_0 = 11.29\pm0.01; 18.535\pm0.001(stat)\pm0.084(syst); 18.917\pm0.002(stat)\pm0.084(syst)$ for M4, the LMC, and the SMC, respectively. Therefore, we conclude that neglecting the metallicity uncertainty does not significantly impact the final results. 

\begin{table}
    \scriptsize
    \centering
    \caption{Bayesian Fitting Results}
    \label{tab:bayesian}
    \begin{tabular}{l c cccc}
    \hline 
    Filters & $\alpha^a$ & a & b & c & $\sigma$ \\
    \hline
    \multicolumn{6}{c}{\emph{PLZ Relations$^b$}} \\
    $I$ & & $ 0.17\pm0.01$ & $-1.43\pm0.17$ & $ 0.28\pm0.03$ & 0.19 \\
    $J$ & & $-0.14\pm0.01$ & $-1.91\pm0.16$ & $ 0.25\pm0.03$ & 0.19 \\
    $H$ & & $-0.31\pm0.01$ & $-2.38\pm0.15$ & $ 0.21\pm0.02$ & 0.18 \\
    $K$ & & $-0.37\pm0.01$ & $-2.43\pm0.15$ & $ 0.21\pm0.02$ & 0.18 \\
    $[3.6]$ & & $-0.39\pm0.01$ & $-2.44\pm0.14$ & $ 0.22\pm0.02$ & 0.18 \\
    $[4.5]$ & & $-0.41\pm0.01$ & $-2.48\pm0.15$ & $ 0.22\pm0.02$ & 0.18 \\
    \multicolumn{6}{c}{\emph{Two-band PWZ Relations$^b$}} \\
    $B,U-B$ & $ 6.228$ & $ 0.30\pm0.03$ & $ 1.71\pm0.43$ & $ 0.23\pm0.06$ & $0.36$ \\
    $V,B-V$ & $ 3.058$ & $-0.27\pm0.01$ & $-2.71\pm0.16$ & $ 0.08\pm0.03$ & $0.18$ \\
    $R,B-R$ & $ 1.689$ & $-0.41\pm0.01$ & $-2.93\pm0.19$ & $ 0.12\pm0.03$ & $0.18$ \\
    $I,V-I$ & $ 1.467$ & $-0.42\pm0.01$ & $-2.56\pm0.16$ & $ 0.19\pm0.03$ & $0.18$ \\
    $J,V-J$ & $ 0.399$ & $-0.43\pm0.01$ & $-2.42\pm0.18$ & $ 0.25\pm0.03$ & $0.20$ \\
    $H,J-H$ & $ 1.618$ & $-0.59\pm0.01$ & $-3.21\pm0.20$ & $ 0.16\pm0.03$ & $0.19$ \\
    $[3.6],V-[3.6]$ & $ 0.071$ & $-0.46\pm0.01$ & $-2.59\pm0.16$ & $ 0.23\pm0.03$ & $0.18$ \\
    $[4.5],K-[4.5]$ & $ 0.918$ & $-0.45\pm0.01$ & $-2.59\pm0.17$ & $ 0.24\pm0.03$ & $0.18$ \\
    \multicolumn{6}{c}{\emph{Three-band PWZ Relations$^b$}} \\
    $B,U-R         $ & $ 1.878$ & $-0.20\pm0.02$ & $-1.48\pm0.20$ & $ 0.16\pm0.03$ & $0.21$ \\
    $V,B-I         $ & $ 1.365$ & $-0.35\pm0.01$ & $-2.54\pm0.14$ & $ 0.14\pm0.02$ & $0.18$ \\
    $R,B-I         $ & $ 1.138$ & $-0.39\pm0.01$ & $-2.64\pm0.15$ & $ 0.14\pm0.02$ & $0.18$ \\
    $I,V-K         $ & $ 0.673$ & $-0.46\pm0.01$ & $-2.62\pm0.14$ & $ 0.19\pm0.02$ & $0.18$ \\
    $J,V-[3.6]     $ & $ 0.306$ & $-0.44\pm0.01$ & $-2.48\pm0.15$ & $ 0.21\pm0.02$ & $0.19$ \\
    $K,I-[3.6]     $ & $ 0.220$ & $-0.49\pm0.01$ & $-2.67\pm0.15$ & $ 0.20\pm0.02$ & $0.18$ \\
    $[3.6],I-[4.5] $ & $ 0.123$ & $-0.46\pm0.01$ & $-2.58\pm0.14$ & $ 0.21\pm0.02$ & $0.18$ \\
    $H,J-K         $ & $ 1.041$ & $-0.55\pm0.01$ & $-2.99\pm0.16$ & $ 0.17\pm0.02$ & $0.18$ \\
    \hline
    \multicolumn{5}{p{.9\linewidth}}{NOTE - Only select relations are shown here, the full version of this table is available online.}\\
    \multicolumn{5}{p{.9\linewidth}}{$^a$ Color coefficient for the Wesenheit magnitude} \\
    \multicolumn{5}{p{.9\linewidth}}{$^b$ $M = a + b(\log P_F +0.3) + c([Fe/H]+1.36)$}
    \end{tabular}
\end{table}


\bsp	
\label{lastpage}
\end{document}